\documentclass[aps,prc,twocolumn,fleqn,floatfix,twoside,superscriptaddress]{revtex4}

\usepackage[dvipdfm]{graphicx}% Include figure files
\usepackage{amsmath,amssymb,amsfonts}

\usepackage{color}

\begin{document}

\title{Radiative jet energy loss in a three-dimensional  hydrodynamical medium
and high $p_T$ azimuthal asymmetry of $\pi_0$ suppression at mid and forward rapidity at RHIC }

\author{Guang-You Qin}
\author{J\"org Ruppert}
\author{Simon Turbide}
\author{Charles Gale}
\affiliation{Department of Physics, McGill University, Montreal, Quebec, H3A 2T8, Canada}

\author{Chiho Nonaka}
\affiliation{Department of Physics, Nagoya University, Nagoya 464-8602, Japan}

\author{Steffen A. Bass}
\affiliation{Department of Physics, Duke University, Durham, NC 27708, USA}

\date{\today}
%%%%%%%%%%%%%%%%%%%%%%%%%%%%%%%%%%%%%%%%%%%%%%%%%%%%%%%%%%%%%%%%%%%%%%%%%%%%%%%%%%%%%%%%%%%%%%%%%%%%%%%%%%%%%%%%%%%%%%%%
\begin{abstract}

The nuclear modification factor $R_{AA}$ for $\pi_0$ production in  Au+Au collisions at $\sqrt{s}=200$~AGeV is
calculated, and studied at high transverse momenta $p_T$. The soft thermalized nuclear medium is described
within the framework of relativistic ideal three-dimensional hydrodynamics. The energy loss of partonic jets is
evaluated in the context of gluon bremsstrahlung in the thermalized partonic matter. We provide a systematic
analysis of  the azimuthal asymmetry of $\pi_0$ suppression at high $p_T$ in central and non-central collisions,
at mid and forward rapidity. The determination of $R_{AA}$ as a function of $p_T$, at different azimuthal
angles, and different rapidities makes for a stringent test of our theoretical understanding of jet energy loss
over a variety of in-medium path lengths, temperatures and initial partonic jet energies. This lays the
groundwork for a tomography of the nuclear medium.
\end{abstract}
\maketitle
%%%%%%%%%%%%%%%%%%%%%%%%%%%%%%%%%%%%%%%%%%%%%%%%%%%%%%%%%%%%%%%%%%%%%%%%%%%%%%%%%%%%%%%%%%%%%%%%%%%%%%%%%%%%%%%%%%%%%%%%
\section{Introduction}

Experiments at the Relativistic Heavy Ion Collider (RHIC) have shown that high $p_T$ hadrons in central A+A
collisions are significantly suppressed in comparison with those in binary p+p interactions, scaled to
nucleus-nucleus collisions \cite{Adcox:2001jp,Adler:2002xw}. This phenomenon is commonly attributed to the fact
that partonic jets produced in the early pre-equilibrium stage of the collisions interact with the hot and dense
nuclear medium created in those collisions and loose energy in the process. This is referred to as {\em
jet-quenching} \cite{Gyulassy:1993hr}. These lower energy partonic jets traverse the medium and will eventually
fragment into hadrons which are observed in the detectors.

Theoretical formalisms have been elaborated to describe the energy loss following the gluon bremsstrahlung
experienced by the color charges in the medium: we mention the work by Baier-Dokshitzer-Mueller-Peigne-Schiff
(BDMPS) \cite{Baier:1996kr}, Gyulassy-Levai-Vitev (GLV) \cite{Gyulassy:2000er}, Kovner-Wiedemann (KW)
\cite{Kovner:2003zj}, Zakharov \cite{Zakharov:1996fv}, Majumder-Wang-Wang (Higher Twist) \cite{Wang:2001if}, and
Arnold-Moore-Yaffe (AMY) \cite{Arnold:2001ms}.

Recent studies \cite{Adil:2006ei,Dutt-Mazumder:2004xk, Mustafa:2003vh} indicate that additional collisional energy loss of light
partons might be substantial as well. However, a consistent treatment including (possibly destructive)
interference is to be developed \cite{Wang:2006qr}. In this article we will restrict ourselves to the
calculation of energy loss of the hard partons induced by gluon bremsstrahlung in the deconfined phase.

Jet quenching can be experimentally quantified by measurements of various quantities as e.g. the nuclear
modification factor $R_{AA}$, the elliptic flow $v_2$ at high $p_T$, and high $p_T$ hadron correlations. While
considerable theoretical effort has been deployed to develop and improve our understanding of modifications of
jets in the nuclear medium, early jet quenching calculations often relied on an elementary description of the
soft medium in their description of data. In most works the jets traverse a simple density distribution which
varies with time unconstrained by the bulk observables, with or without a Bjorken expansion, see e.g.
\cite{Dainese:2004te,Majumder:2006we}. Similarly, calculations estimating the effects of three dimensional (3D)
expansion on $R_{AA}$ have treated the energy loss of jets in a simplified fashion \cite{Hirano:2003hy}.

In \cite{Renk:2005ta} a parameterized non-Bjorken fireball evolution that accounts for several measured
observables connected with bulk properties of the matter created at RHIC \cite{Renk:2004yv} was applied to study
the effect of flow on energy loss (in the BDMPS formalism, according to the prescription outlined in
\cite{Salgado:2003gb}). This study aimed for a sophisticated description of energy loss as well as for the
medium evolution, but it was restricted to the calculation of $R_{AA}$ in central collisions. Later this
observable was also studied in a 2D hydrodynamical evolution model \cite{RenkEskola}. Recently, a 3D
hydrodynamical evolution calculation \cite{Nonaka:2006yn} of the expanding medium in central and non-central
collisions was employed in detailed studies of jet energy loss as predicted in the BDMPS formalism
\cite{Renk:2006sx,Bass:2007em} and in the higher twist formalism \cite{Majumder2}.

The present work contributes to this effort of understanding the physics of jet quenching by applying the
Arnold, Moore, and Yaffe (AMY) formalism \cite{Arnold:2001ms} for gluon bremsstrahlung to calculate the jet
energy loss in the thermal partonic medium in central and non-central collision as inferred from 3D relativistic
hydrodynamics \cite{Nonaka:2006yn}. We present a calculation of $R_{AA}$ as a function of transverse momentum
(and the azimuth) in central and non-central collisions and also study the rapidity dependence of this quantity.

While $R_{AA}$ as measured in central collisions \emph{alone} is not suited to distinguish in detail between
different theoretical conjectures about jet energy loss \cite{Renk:2006pw}, the combination with additional
measurements  of $R_{AA}$ versus reaction plane in non-central collisions \cite{Adler:2} and at finite rapidity
provides further valuable tomographic information. Additional tomographic observables are high $p_t$ triggered
correlation measurements, see e.g. \cite{RenkEskola,RenkRuppert1,RenkRuppert2}.

The paper is organized as follows, we first briefly review the 3D hydrodynamical description of the medium in Sec. II.  We then discuss in Sec. III how the initial momentum distributions of jets and their time-evolution in the thermal medium (which incorporates the energy loss process in the AMY formalism) as well as the
fragmentation of the final jets into pions are calculated. Numerical results are presented for $R_{AA}$ at mid
and forward rapidity in Sec. IV together with a comparison to data where already available.
Finally, Sec. V
contains our conclusions.

%%%%%%%%%%%%%%%%%%%%%%%%%%%%%%%%%%%%%%%%%%%%%%%%%%%%%%%%%%%%%%%%%%%%%%%%%%%%%%%%%%%%%%%%%%%%%%%%%%%%%%%%%%%%%%%%%%%%%%%%
\section{3D hydrodynamical medium}
%%%%%%%%%%%%%%%%%%%%%%%%%%%%%%%%%%%%%%%%%%%%%%%%%%%%%%%%%%%%%%%%%%%%%%%%%%%%%%%%%%%%%%%%%%%%%%%%%%%%%%%%%%%%%%%%%%%%%%%%

The behavior related to the bulk properties of the high-density phase in heavy-ion collisions at RHIC is well
described by Relativistic Fluid Dynamics (RFD, see e.g. \cite{Bjorken:1982qr,Clare:1986qj,Dumitru:1998es}),
while this description is not applicable in the late dilute stages of the collisions in which the mean free path
of hadrons is large on the typical scales of the system.

In the present paper we use a fully 3D hydrodynamical model for the description of RHIC physics
\cite{Nonaka:2006yn} which solves the relativistic hydrodynamical equation
\begin{eqnarray}
\partial_\mu T^{\mu \nu} = 0,
\label{Eq-rhydro}
\end{eqnarray}
where $T^{\mu \nu}$ is the energy momentum tensor which can be expressed as
\begin{eqnarray}
T^{\mu \nu}=(\epsilon + p) U^{\mu} U^{\nu} - p g^{\mu \nu}.
\end{eqnarray}
Here $\epsilon$, $p$, $U$ and $g^{\mu \nu}$ are energy density, pressure, four velocity and metric tensor,
respectively. Furthermore baryon number $n_B$ conservation is imposed as a constraint
\begin{eqnarray}
\partial_\mu (n_B (T,\mu) U^\mu)=0,
\end{eqnarray}
and the resulting set of partial differential equations is closed by specifying an equation of state (EoS):
$\epsilon = \epsilon(p)$. Our particular RFD calculation utilizes a Lagrangian mesh and the coordinates
$(\tau,x,y,\eta)$ with the longitudinal proper time $\tau=\sqrt{t^2-z^2}$ and space-time rapidity
$\eta=\frac{1}{2}\ln[(t+z)/(t-z)]$ in order to optimize the calculation for the ultra-relativistic regime of
heavy collisions at RHIC. Once an initial condition has been specified RFD in the ideal fluid approximation (i.
e. neglecting off-equilibrium effects) allows a calculation of single soft matter properties at RHIC, especially
collective flow and particle spectra.

We assume early thermalization with subsequent hydrodynamical expansion at $\tau_0=0.6~{\rm fm/c}$. The initial
conditions, namely initial energy density and baryon number density are parameterized by
\begin{eqnarray}
\epsilon(x,y,\eta)& =& \epsilon_{\rm max}W(x,y;b)H(\eta),
\nonumber \\
n_B(x,y,\eta)& = & n_{B{\rm max}}W(x,y;b)H(\eta),
\end{eqnarray}
where $b$ and $\epsilon_{\rm max}$ ($n_{B{\rm max}}$) are the impact parameter and the maximum value of energy
density (baryon number density), respectively. $W(x,y;b)$ is given by a combination of wounded nucleon model and
binary collision model \cite{Kolb:2001qz} and $H(\eta)$ is given by $\displaystyle H(\eta)=\exp \left [ -
(|\eta|-\eta_0)^2/(2 \sigma_\eta^2) \cdot \theta ( |\eta| - \eta_0 ) \right ]$.

The initial conditions have been chosen such that a successful description of the soft sector at RHIC (elliptic
flow, pseudo-rapidity distributions and low-$p_T$ single particle spectra) is achieved. For further details,
especially also a discussion of the EoS which is employed, we refer the reader to \cite{Nonaka:2006yn}.

%%%%%%%%%%%%%%%%%%%%%%%%%%%%%%%%%%%%%%%%%%%%%%%%%%%%%%%%%%%%%%%%%%%%%%%%%%%%%%%%%%%%%%%%%%%%%%%%%%%%%%%%%%%%%%%%%%%%%%%%
\section{Jet evolution and fragmentation}
%%%%%%%%%%%%%%%%%%%%%%%%%%%%%%%%%%%%%%%%%%%%%%%%%%%%%%%%%%%%%%%%%%%%%%%%%%%%%%%%%%%%%%%%%%%%%%%%%%%%%%%%%%%%%%%%%%%%%%%%

In this section we present the techniques used to calculate the initial jet production in the early stage of the
collisions, the subsequent propagation through the hot and dense medium, and final hadronization in the vacuum.
We exclusively focus on the hadrons in the high $p_T$ region in which fragmentation is the dominant mechanism
for the production of hadrons. For softer hadrons (below $p_T \sim 7$~GeV/c) other mechanisms, such as the
recombination of partons  become of increasing significance \cite{Fries:2003kq}.

The initial jet density distribution $\mathcal{P}_{AB}(b,\vec{r}\bot)$ at the transverse position $\vec{r}_\bot$
and in A+B collisions with  impact parameter $\vec{b}$ is given by
\begin{eqnarray}
\mathcal{P}_{AB}(b,\vec{r}_\bot) &=& \frac{T_A(\vec{r}_\bot + \vec{b}/2)T_B(\vec{r}_\bot -
\vec{b}/2)}{T_{AB}(b)}.
\end{eqnarray}
Here we use a Woods-Saxon form for the nuclear density function,
$\rho(\vec{r}_\bot,z)={\rho_0}/[{1+\exp(\frac{r-R}{d})}]$, to evaluate the nuclear thickness function
$T_A(\vec{r}_\bot)=\int dz \rho_A(\vec{r}_\bot,z)$ and the overlap function of two nuclei $T_{AB}(b)=\int
    d^2r_\bot T_A(\vec{r}_\bot) T_B(\vec{r}_\bot+\vec{b})$. {The values of the parameters  $R=6.38~{\rm fm}$ and $d=0.535~{\rm fm}$ are taken from \cite{DeJager:1974dg}.}

The initial momentum distribution ${dN^{j}_{AB}(b)}/{d^2p^j_Tdy}|_{i}$ of jets  is computed from pQCD
 in the factorization formalism,
\begin{eqnarray}
\label{pdfeqn}
 \left.\frac{dN^{j}_{AB}(b)}{d^2p^j_Tdy}\right|_{i} &=& T_{AB}(b) \sum_{abd} \int dx_a
G_{a/A}(x_a,Q)G_{b/B}(x_b,Q) \nonumber \\ && \times \frac{1}{\pi} \frac{2x_ax_b}{2x_a-x^j_Te^y} K
\frac{d\sigma_{a+b\to j+d}}{dt}.
\end{eqnarray}
{In the above equation, $G_{a/A}(x_a,Q)$ is the distribution function of parton $a$ with momentum fraction $x_a$
in the nucleus $A$ at factorization scale $Q$, taken from CTEQ5 \cite{Lai:1999wy} including nuclear shadowing
effects {from EKS98} \cite{Eskola:1998df}. The index $j$ represents one of the partonic species
($j=q,\bar{q},g$), and $x_T^j = 2p_T^j/\sqrt{s_{NN}}$,
where $\sqrt{s_{NN}}$ is the center of mass energy. The distribution ${d\sigma}/{dt}$
is the leading order QCD differential cross section, and the $K$-factor accounts for NLO effects and is taken to
be constant in our calculation as it is almost $p_T$ independent \cite{Eskola:2005ue, Jager:2002xm,
Barnafoldi:2000dy}.} The initial Cronin effect is neglected in our calculation since the nuclear modification
factor of neutral pions from d+Au collisions measured by PHENIX is consistent with $1$ within systematic errors
\cite{Adler:2006wg}.

The evolution of a jet momentum distribution $P_j(p,t)={dN_j(p,t)}/{dpdy}$ (essentially the probability of
finding a jet with energy $p$ at time $t$) in the medium is obtained in the AMY formalism by solving a set of
coupled rate equations (for details see \cite{Jeon:2003gi, Turbide:2005fk}), which have the following generic
form,
\begin{eqnarray}\label{jet-evolution-eq}
\frac{dP_j(p,t)}{dt} &=& \sum_{ab} \int dk \left[P_a(p+k,t) \frac{d\Gamma^{a}_{jb}(p+k,k)}{dk dt}
\right.\nonumber\\ && \left. - P_j(p,t)\frac{d\Gamma^{j}_{ab}(p,k)}{dk dt}\right],
\end{eqnarray}
where ${d\Gamma^{j}_{ab}(p,k)}/{dk dt}$ is the transition rate for the partonic process $j\to a+b$. We point out
that the calculation includes not only the emission but also the absorption of thermal partons as the $k$
integral in Eq.~(\ref{jet-evolution-eq}) ranges from $-\infty$ to $\infty$. The transition rate is given by
\cite{Jeon:2003gi, Turbide:2005fk}
\begin{eqnarray}\label{eq:dGamma}
\frac{d\Gamma(p,k)}{dk dt} & = & \frac{C_s g_s^2}{16\pi p^7}
        \frac{1}{1 \pm e^{-k/T}} \frac{1}{1 \pm e^{-(p-k)/T}}
\nonumber \\ && \times \left\{ \begin{array}{cc}
        \frac{1+(1{-}x)^2}{x^3(1{-}x)^2} & q \rightarrow qg \\
        N_{\rm f} \frac{x^2+(1{-}x)^2}{x^2(1{-}x)^2} & g \rightarrow q\bar{q} \\
        \frac{1+x^4+(1{-}x)^4}{x^3(1{-}x)^3} & g \rightarrow gg \\
        \end{array} \right\}
\nonumber \\ && \times \int \frac{d^2 \vec{h}}{(2\pi)^2} 2 \vec{h} \cdot {\rm Re}\: \vec{F}(\vec{h},p,k) \, .
\end{eqnarray}
Here $C_s$ is the quadratic Casimir relevant for the process, and $x\equiv k/p$ is the momentum fraction of the
gluon (or the quark, for the case $g \rightarrow q\bar{q}$). $\vec{h} \equiv \vec{p} \times \vec{k}$ determines
how non-collinear the final state is; it is treated as parametrically $O(g_sT^2)$ and therefore small compared
to $\vec{p}\cdot \vec{k}$. Therefore it can be taken as a two-dimensional vector in transverse space.
$\vec{F}(\vec{h},p,k)$ is the solution of the following integral equation~\cite{Jeon:2003gi, Turbide:2005fk}:
\begin{eqnarray}\label{eq:integral_eq1}
2\vec{h} &=&
        i \delta E(\vec{h},p,k) \vec{F}(\vec{h}) + g_s^2 \int \frac{d^2 \vec{q}_\perp}{(2\pi)^2}
C(\vec{q}_\perp) \nonumber \\ && \times
   \Big\{ (C_s-C_{\rm A}/2)[\vec{F}(\vec{h})-\vec{F}(\vec{h}{-}k\,\vec{q}_\perp)]
        \nonumber \\ &&
        + (C_{\rm A}/2)[\vec{F}(\vec{h})-\vec{F}(\vec{h}{+}p\,\vec{q}_\perp)]
        \nonumber \\ &&
        +(C_{\rm A}/2)[\vec{F}(\vec{h})-\vec{F}(\vec{h}{-}(p{-}k)\,\vec{q}_\perp)] \Big\} \, .
\end{eqnarray}
Here $\delta E(\vec{h},p,k)$ is the energy difference between the final and the initial states:
%\cite{Jeon:2003gi, Turbide:2005fk},
\begin{eqnarray}\label{eq:integral_eq2}
\delta E(\vec{h},p,k) &=& \frac{\vec{h}^2}{2pk(p{-}k)} + \frac{m_k^2}{2k} + \frac{m_{p{-}k}^2}{2(p{-}k)} -
\frac{m_p^2}{2p}\, , \ \ \ \ \ \
\end{eqnarray}
and $m^2$ are the medium induced thermal masses. Also, $C(\vec{q}_\perp)$ is the differential rate to exchange
transverse (to the parton) momentum $\vec{q}_\perp$. In a hot thermal medium, its value at leading order in
$\alpha_{\rm s}$ is \cite{Aurenche:2002pd}
\begin{eqnarray}\label{eq:Cq}
C(\vec{q}_\perp) = \frac{m_D^2}{\vec{q}_\perp^2(\vec{q}_\perp^2{+}m_D^2)} \, , \ \ \ m_D^2 = \frac{g_s^2 T^2}{6}
(2 N_{\rm c} {+} N_{\rm f})\, . \ \ \
\end{eqnarray}
For the case of $g\rightarrow q\bar{q}$, $(C_s-C_{\rm A}/2)$ should appear as the prefactor on the term
containing $\vec{F}(\vec{h}-p\,\vec{q}_\perp)$ rather than $\vec{F}(\vec{h}-k \,\vec{q}_\perp)$.

The strength of the transition rate in pQCD is controlled by the strong coupling constant $\alpha_s$,
temperature $T$ and the flow parameter $\vec{\beta}$ (the velocity of the thermal medium) relative to the jet's
path. In a 3D expanding medium, the transition rate is first evaluated in the local frame of the thermal medium,
then boosted into the laboratory frame,
\begin{eqnarray}
\label{eqflow} \left.\frac{d\Gamma(p,k)}{dk dt}\right|_{lab} &=& (1-\vec{v}_j\cdot\vec{\beta})
\left.\frac{d\Gamma(p_0,k_0)}{dk_0dt_0}\right|_{local}, \ \ \ \ \ \
\end{eqnarray}
where $k_0=k(1-\vec{v}_j\cdot\vec{\beta})/\sqrt{1-\beta^2}$ and $t_0=t\sqrt{1-\beta^2}$ are momentum and the
proper time in the local frame, and $\vec{v}_j$ is the velocity of the jet. As jets propagate in the medium, the
temperature and the flow parameter depend on the time and the positions of jets, and the 3D hydrodynamical
calculation \cite{Nonaka:2006yn} is utilized to determine the temperature and flow profiles. The energy-loss
mechanism is applied at $\tau_0=0.6$~fm/c, when the medium reaches thermal equilibrium, and turned off when the
medium reaches the hadronic phase.

\begin{figure}[htb]
\begin{center}
\includegraphics[width=8cm]{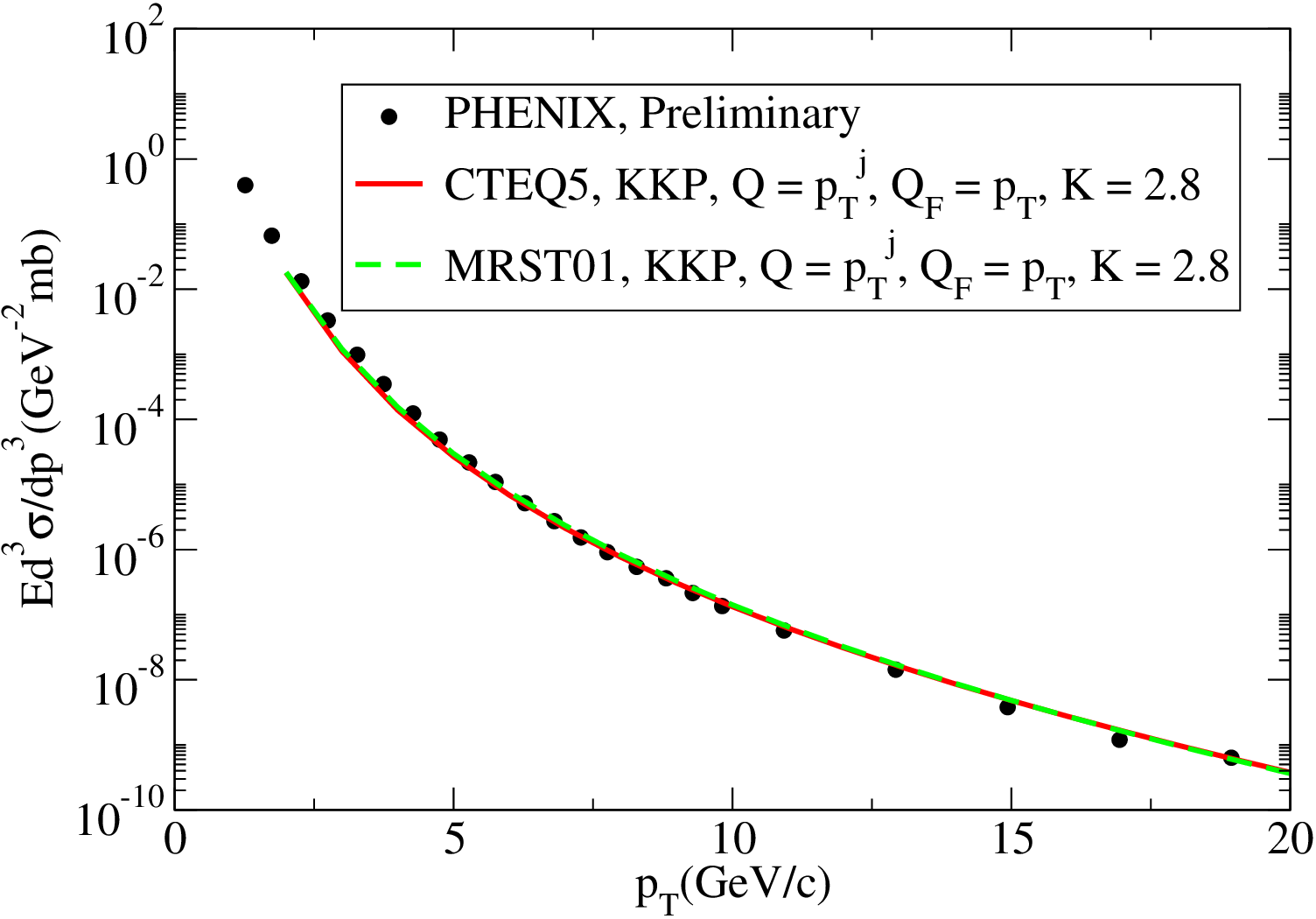}
\end{center}
\caption{(Color online) {The inclusive cross section for $\pi_0$ production versus $\pi_0$ transverse momentum
at mid-rapidity in pp collisions at $\sqrt{s}=200~{\rm GeV}$, compared with PHENIX data \cite{phenix-pi0-pp}.}
}\label{pp-cross-section}
\end{figure}

\begin{figure}[htb]
\begin{center}
\includegraphics[width=8cm]{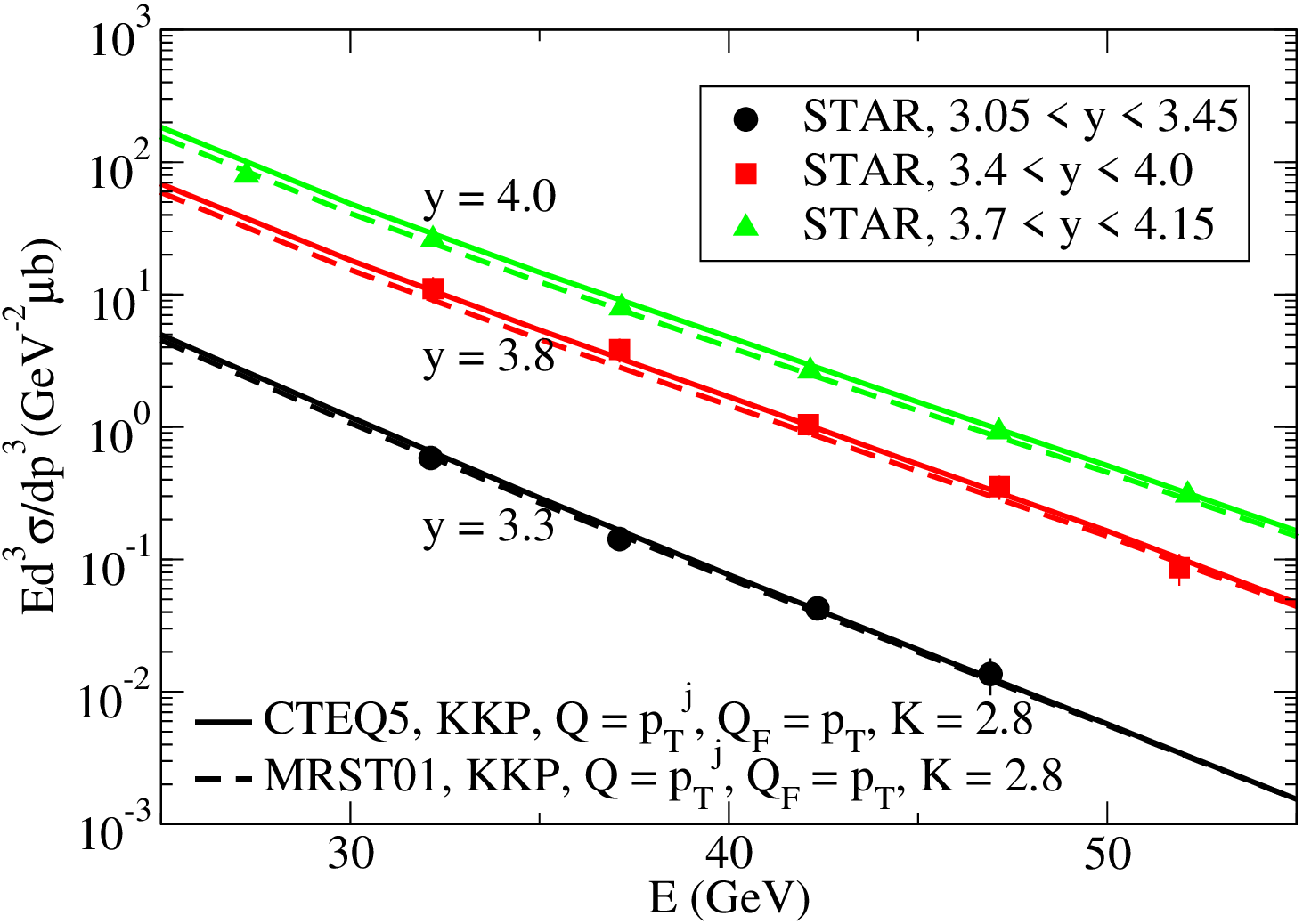}
\end{center}
\caption{(Color online) {The inclusive cross section for $\pi_0$ production versus $\pi_0$ energy at forward
rapidity in pp collisions at $\sqrt{s}=200~{\rm GeV}$. Data points are taken from STAR \cite{Adams:2006uz}.}
}\label{pp-cross-section-y}
\end{figure}

The final hadron spectrum ${dN^h_{AB}(b)}/{d^2p_Tdy}$ at high $p_T$ is obtained by the fragmentation of jets in
the vacuum after their passing through the 3D expanding medium,
\begin{eqnarray}
\label{hadron_distr} \frac{dN^h_{AB}(b)}{d^2p_Tdy} &=& \sum_{j} \int d^2\vec{r}_\bot
\mathcal{P}_{AB}(b,\vec{r}_\bot) \int \frac{ dz_j}{z_j^2} {D_{h/j}(z_j,Q_F)} \nonumber \\ && \times \left.
\frac{dN^{j}_{AB}(b,\vec{r}_\bot)}{d^2p^j_Tdy}\right|_{f},
\end{eqnarray}
where ${dN^{j}_{AB}(b,\vec{r}_\bot)}/{d^2p^j_Tdy}|_{f}$ is the final momentum distribution of the jet initially
created at transverse position $\vec{r}_\bot$ after passing through the medium. {The fragmentation function
${D_{h/j}(z_j,Q_F)}$ gives the average multiplicity of the hadron $h$ with the momentum fraction $z_j=p_T/p^j_T$
produced from a jet $j$ at a scale $Q_F$, taken from KKP parametrization \cite{Kniehl:2000fe}. The
fractorization scale $Q = p_T^j$ and fragmentation scale $Q_F = p_T$ are set as in \cite{Eskola:2005ue} where
the $K$-factor is found to be $2.8$.} {We use these values throughout the present study. This nicely reproduces
the experimentally measured $\pi_0$ yield at mid and forward rapidity in p+p collisions at $\sqrt{s} = 200$~GeV,
as shown in Fig.~\ref{pp-cross-section} and Fig.~\ref{pp-cross-section-y}.
It can be clearly seen that replacing the CTEQ5 parton distribution functions by MRST01 \cite{Martin:2002dr} yields essentially the same result for the inclusive $\pi_0$ production in p+p collisions.}
We point out that the presence of a
nuclear medium might in principle alter these scales but we postpone a detailed study of this possibility to
future research.

The nuclear modification factor $R_{AA}$ is defined as the ratio of the hadron yield in A+A collisions to that
in p+p interactions scaled by the number of binary collisions
\begin{eqnarray}
R^h_{ AA}(b,\vec{p}_T,y) &=& \frac{1}{N_{coll}(b)} \frac{{dN^h_{ AA}}(b)/{d^2p_Tdy}} {{dN^h_{ pp}}/{d^2p_Tdy}}.
\end{eqnarray}

%%%%%%%%%%%%%%%%%%%%%%%%%%%%%%%%%%%%%%%%%%%%%%%%%%%%%%%%%%%%%%%%%%%%%%%%%%%%%%%%%%%%%%%%%%%%%%%%%%%%%%%%%%%%%%%%%%%%%%%%
\section{Results}
%%%%%%%%%%%%%%%%%%%%%%%%%%%%%%%%%%%%%%%%%%%%%%%%%%%%%%%%%%%%%%%%%%%%%%%%%%%%%%%%%%%%%%%%%%%%%%%%%%%%%%%%%%%%%%%%%%%%%%%%

\begin{figure}[htb]
\begin{center}
\includegraphics[width=8cm]{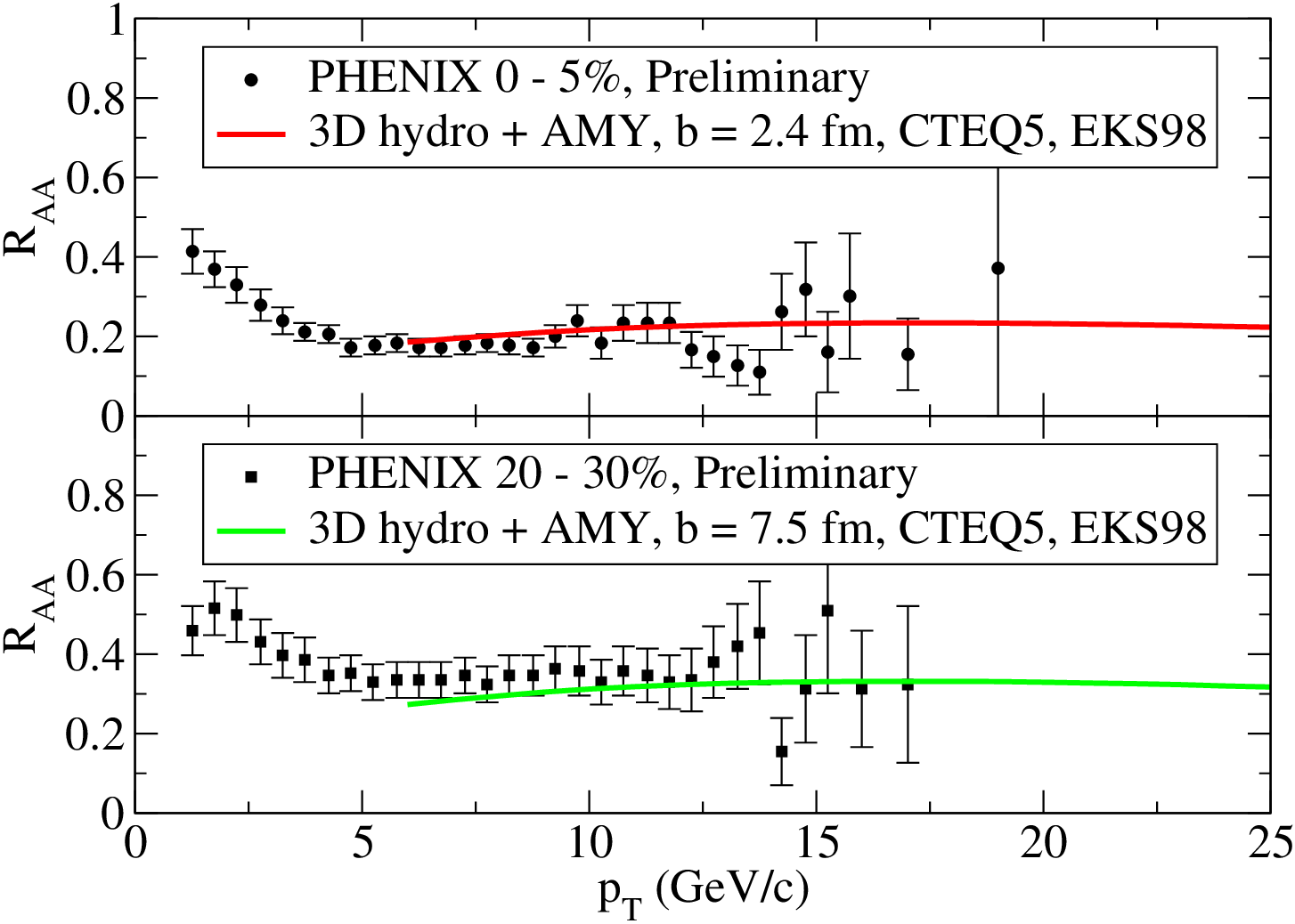}
\end{center}
\caption{(Color online) The neutral pion $R_{AA}$ at midrapidity in most central (upper panel) and midperipheral
(lower panel) Au+Au collisions compared with PHENIX data. }\label{RAA}
\end{figure}

In Fig.~\ref{RAA}, we present the calculation of the nuclear modification factor $R_{AA}$ for neutral pions
measured at midrapidity for two different impact parameters $b=2.4$~fm and $b=7.5$~fm, compared with
(preliminary) PHENIX data for most central ($0-5\%$) and midcentral ($20-30\%$) collisions \cite{Adler:2}. All
results presented throughout the paper are for Au+Au collisions at $\sqrt{s} = 200$~AGeV. {We only show results
for the nuclear modification factor $R_{AA}$ for neutral pions, as results for charged hadrons (including
contributions from charged pions, kaons and protons) are qualitatively similar.}

Once the temperature evolution is fixed by the initial conditions and subsequent 3D hydrodynamical expansion,
the strong coupling constant $\alpha_s$ is the only quantity which is not uniquely determined in the model. The
value of $\alpha_s$ is a direct measure of the interaction strength and is adjusted in such a way that the
experimental data in the most central collisions is described. {The same value, $\alpha_s=0.33$}, is used in
peripheral collisions. Treating $\alpha_s$ as $T$-independent from early thermalization on down to the phase
transition temperature is a simplification and corresponds to the assumption that the deconfined phase of the
medium formed in Au+Au collisions at $200$~AGeV at RHIC can be characterized by one average effective coupling.
We point out that a treatment in the AMY formalism only considers energy loss in the partonic phase, hadronic
energy loss is not included in the present study \footnote{Note that the energy loss in the hadronic medium is
found to be subdominant in \cite{Renk:2006sx}, while data indicate that it has to be assumed to be more
significant in the higher twist formalism \cite{Majumder2}.}. We have verified that choosing different constant
values of $\alpha_s$ does not influence the shape of $R_{AA}$ as a function of $p_T$ significantly while only the overall normalization is affected. We point out that although $\alpha_s<1$, $g_s=\sqrt{4 \pi \alpha_s}$ is actually larger than $1$.
In that sense our study does not contradict the finding in \cite{Renk:2006sx} that a stronger quenching power
of the medium has to be assumed than  if a fully perturbative treatment of jet quenching in the quark gluon plasma is employed. (In \cite{Renk:2006sx} uncertainties in the selection of the strong coupling and possible non-perturbative effects were
parameterized by a factor $K$ in $\hat{q}=2 K \epsilon^{3/4}$. $K \approx 3.6$ was adjusted
to give a good description of $R_{AA}$ in central collisions at mid-rapidity.)

In Fig.~\ref{RAA}, $R_{AA}$ at midrapidity is averaged over the azimuth $\phi$. More tomographic capabilities
can be achieved if one studies $R_{AA}$ at midrapidity in {\em non-central} collisions not only as a function of
$p_T$ averaged over $\phi$ but also as a function of the azimuth $\phi$ \cite{Adler:2}.

The reason is that the initial geometric asymmetry in non-central collisions leaves its imprint on the 3D
hydrodynamical evolution and initial jets experience different energy loss (depending on where they are produced
in the medium and in which direction they are emitted) owing to the different local properties of the nuclear
medium with which they interact. In the AMY formalism the important input from the evolution is the temperature
in the rest frame of the local fluid that the jet experiences (and to a lesser extent the flow profile of the
medium, as discussed later).

\begin{figure}[htb]
\begin{center}
\includegraphics[width=8cm]{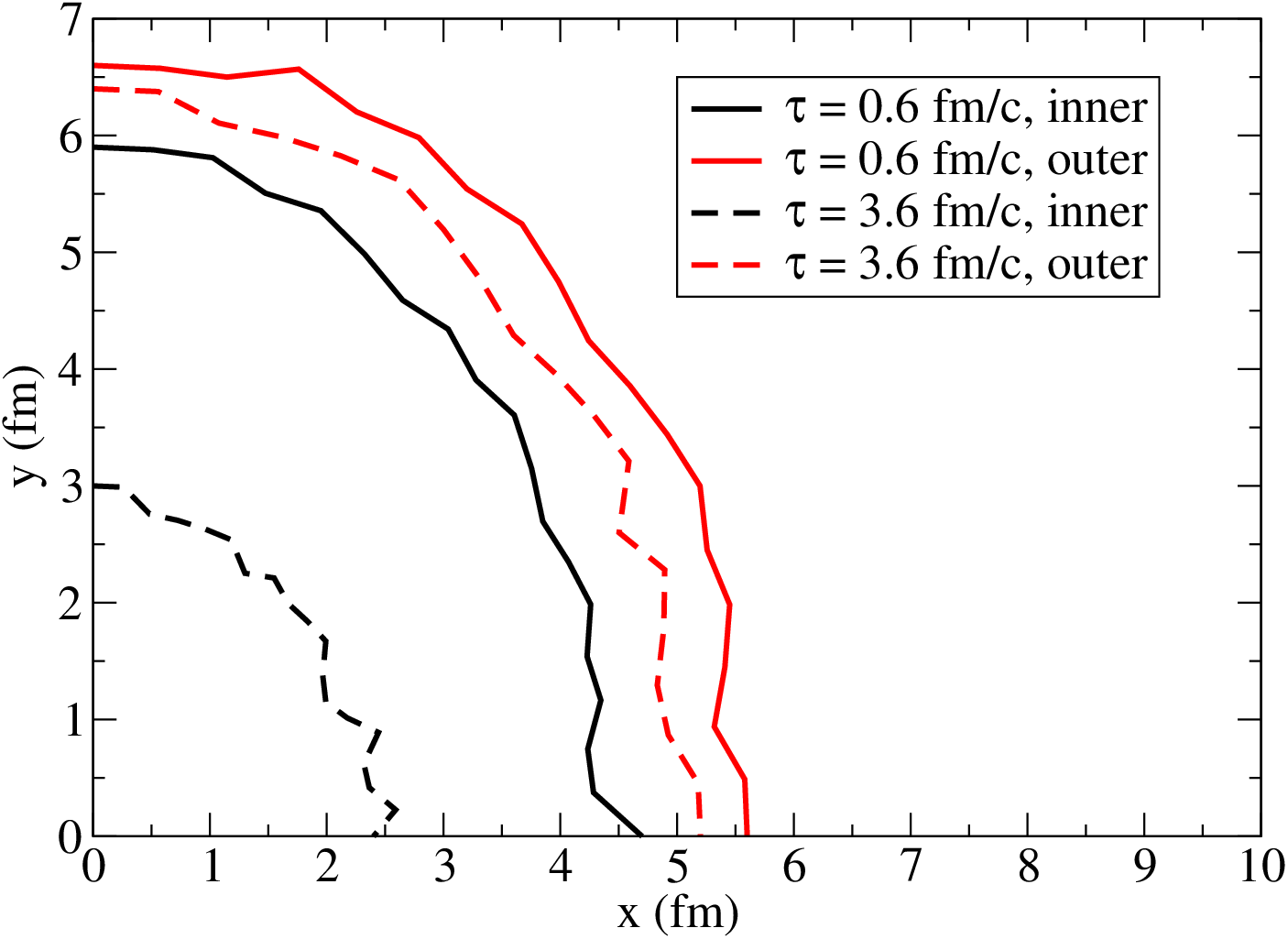}
\end{center}
\caption{(Color online) The inner and outer boundaries for $T=T_c$ in the transverse plane at two different
proper times, $b=7.5$~fm. }\label{boundary}
\end{figure}

\begin{figure}[htb]
\begin{center}
\includegraphics[width=8cm]{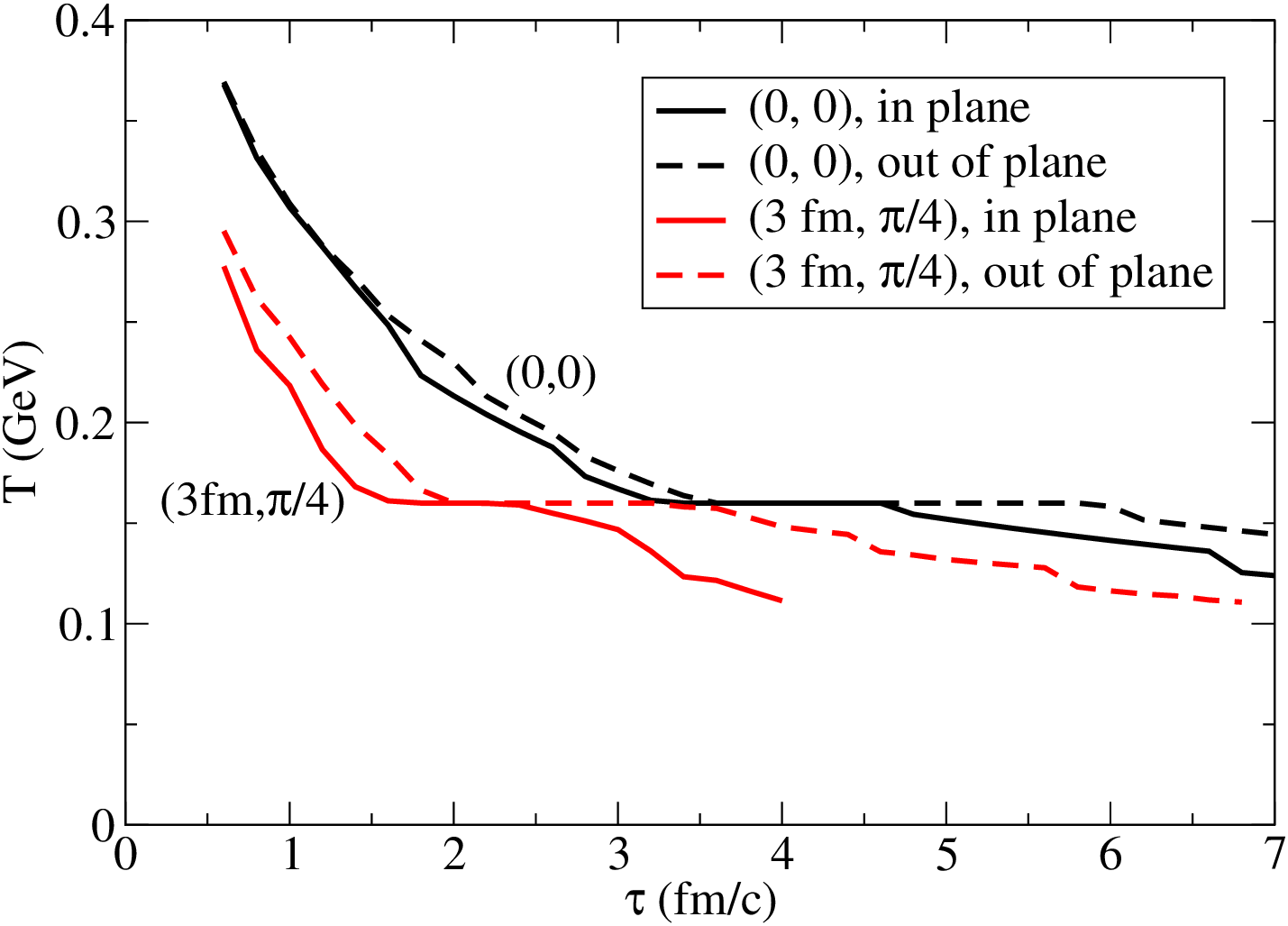}
\end{center}
\caption{(Color online) The time evolution of the temperature seen by a jet initially created at $(r_0,\phi_0)$
moving in plane and out of plane through the medium, $b=7.5$~fm. }\label{jet-temperature}
\end{figure}

To illustrate the geometrical asymmetry we show in Fig.~\ref{boundary} isotherms for $T=T_c$ in the transverse
plane for an impact parameter of $b=7.5$~fm at two different proper times of the evolution. They represent the
inner and outer boundaries of the mixed phase during the evolution. The geometric asymmetry of the temperature
profile can be clearly seen from the plot. Both boundaries move towards the center and the inner boundary moves
faster than the outer boundary. It is useful to define the emission in plane ($\phi=0$) versus out of plane
($\phi=\pi/2$). We point out that the ratios of the boundary positions in plane to those out of plane are almost
constant in proper time and have almost the same values for the inner and outer boundaries: $\sim 0.8$.

Fig.~\ref{jet-temperature} shows the temperature observed by a jet traversing this medium. The jet is assumed to
be created at position $(r_0,\phi_0)$ by a hard scattering at early times in the heavy-ion collision. As it
propagates through the medium, the surrounding environment will change from the QGP phase to the mixed phase,
then to the hadronic phase and will eventually freeze-out. We plot the temperature evolution experienced by jets
that are created in a symmetric position ($\phi_0=\pi/4$) relative to in-plane and out-of-plane and illustrate
the geometrical asymmetry of the medium. We compare jets starting at the origin and those at $r=3$~fm.

Jets that propagate out of plane will pass the mixed phase and the hadronic phase at later proper time than
those traversing in plane and will interact with the deconfined and mixed phase of the medium longer. If the
jets have identical initial energy, the energy loss experienced by the jets propagating out of plane will
therefore be larger than in plane.

\begin{figure}[htb]
\begin{center}
\includegraphics[width=8cm]{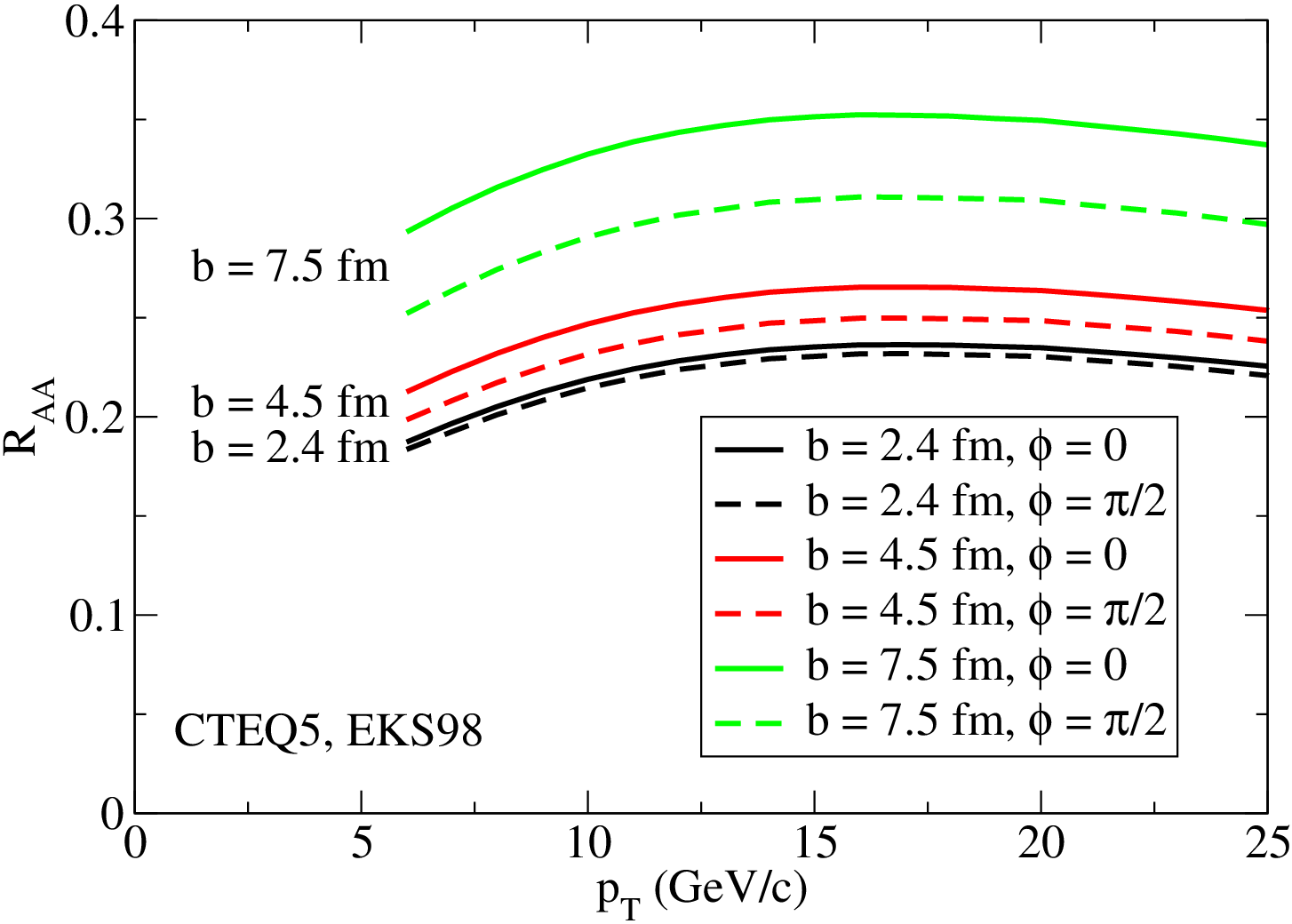}
\end{center}
\caption{(Color online) The neutral pion $R_{AA}$ at midrapidity for emissions in plane and out of plane as a
function of $p_T$ for different impact parameters.}\label{raa-phi}
\end{figure}

\begin{figure}[htb]
\begin{center}
\includegraphics[width=8cm]{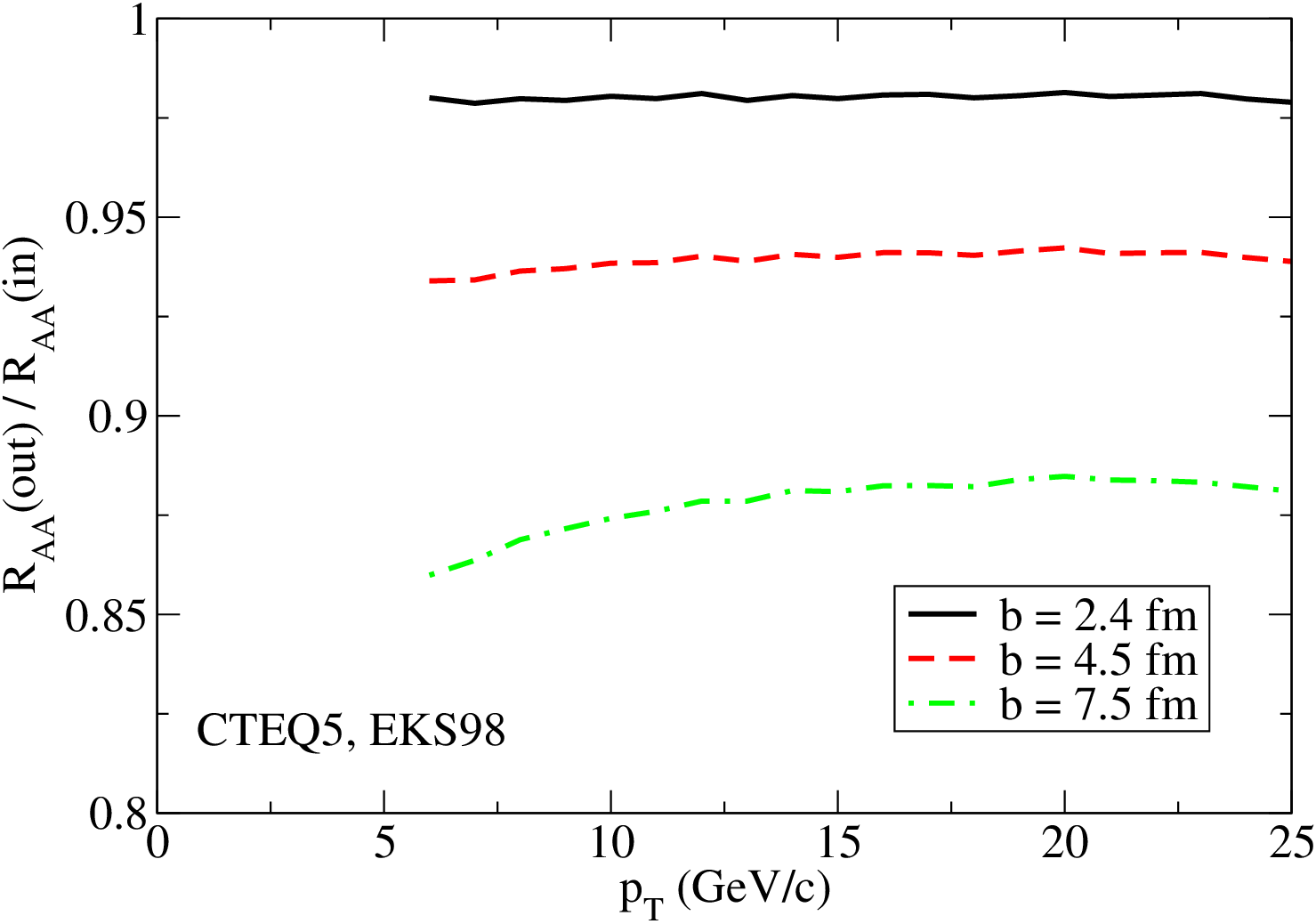}
\end{center}
\caption{(Color online) The ratio of the neutral pion $R_{AA}$ at midrapidity for emissions in plane and out of
plane as a function of $p_T$ for different impact parameters.}\label{raa-phi-ratio}
\end{figure}

This behavior is reflected in $R_{AA}$ as a function of $p_T$ for emissions in plane and out of plane in
Fig.~\ref{raa-phi} for different impact parameters. While there is very small difference for $R_{AA}$ between
the two planes in central collisions, a much larger difference for midcentral collisions {(about 13\% for
$b=7.5$~fm)} is predicted, as can been seen from the ratio of $R_{AA}$ for emission out of plane to that in
plane as shown in Fig.~\ref{raa-phi-ratio}.

\begin{figure}[htb]
\begin{center}
\includegraphics[width=8cm]{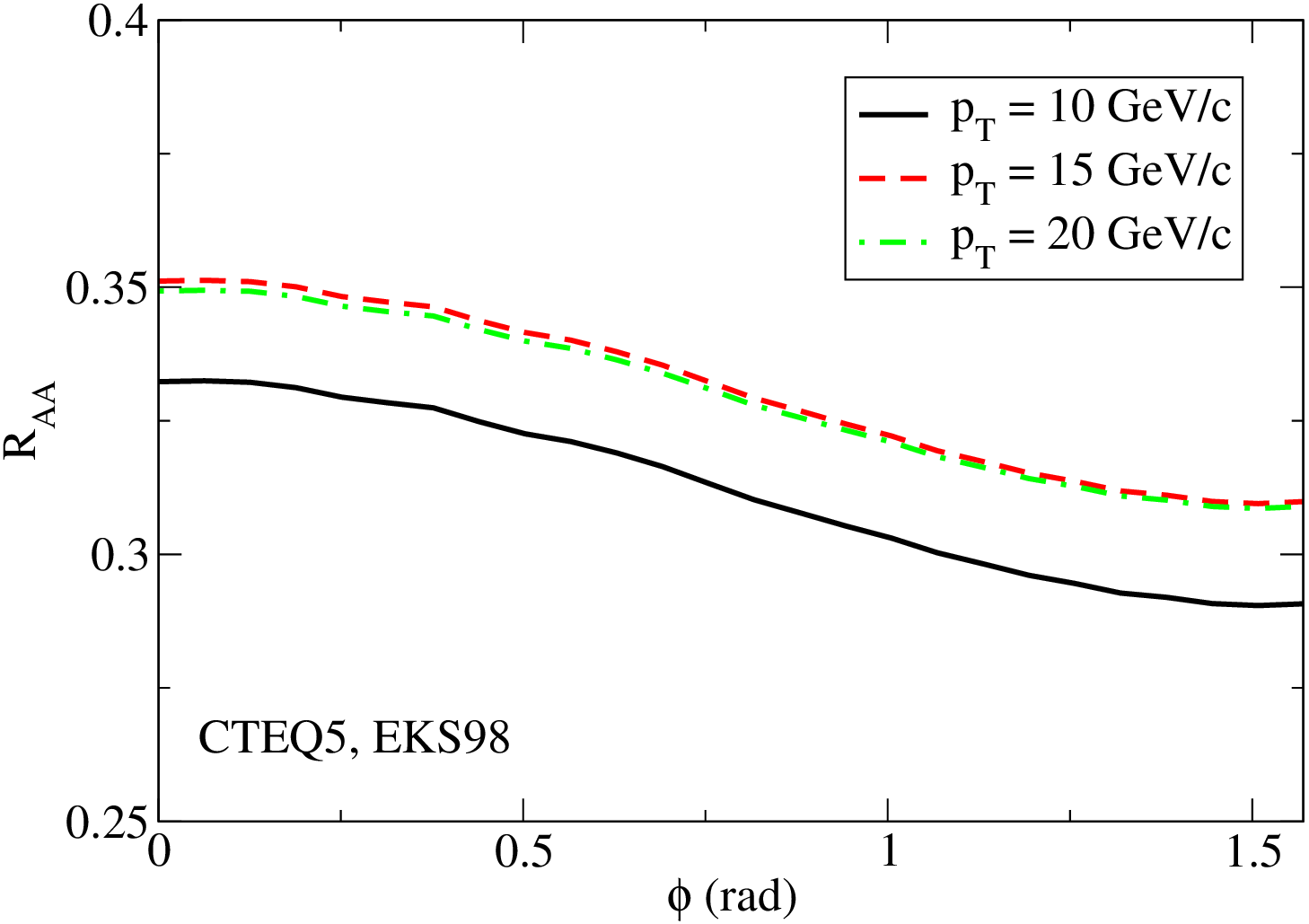}
\end{center}
\caption{(Color online) The neutral pion $R_{AA}$ at midrapidity as a function of the azimuthal angle $\phi$ of
the pion for different $p_T$, $b=7.5$~fm. }\label{raa-pT}
\end{figure}

As a further tomographic quantity, one can also study $R_{AA}$ for non-central collisions as a function of the
azimuthal angle $\phi$ for different $p_T$, see Fig.~\ref{raa-pT}. A monotonous decrease of $R_{AA}$ for
emissions from in plane to out of plane, reflects (an average of) the asymmetric temperature (and flow) profiles
experienced by the jets while they traverse the medium.

\begin{figure}[htb]
\begin{center}
\includegraphics[width=8cm]{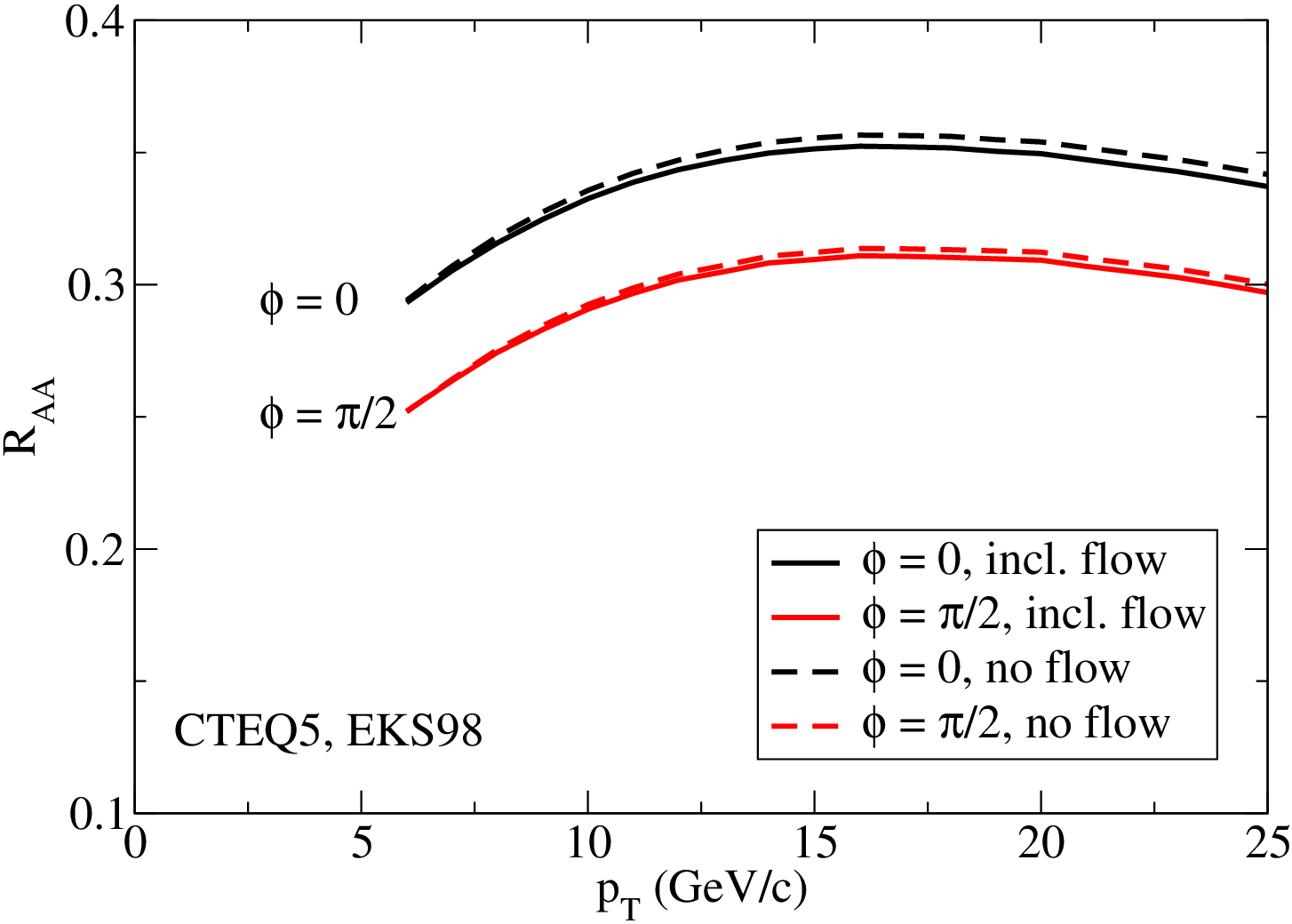}
\end{center}
\caption{(Color online) Comparing the neutral pion $R_{AA}$ at midrapidity with and without flow for emissions
in plane and out of plane as a function of $p_T$, $b=7.5$~fm.}\label{raa-phi-no-flow}
\end{figure}

In a 3D expanding medium, there is also considerable collective flow being built up during the evolution. This
can affect the energy loss of jets and may to some degree influence the asymmetry in the final pion spectrum. To
quantify this effect, we use the same 3D hydro temperature profile, but disregard the transverse flow. We
compare the case with flow to one where the velocity effect is disregarded, namely $\vec{\beta}=0$ is enforced
by hand in Eq.~(\ref{eqflow}) (only for illustration purposes). This treatment can give an estimate on how
collective flow (not the temperature of the medium) influences the jet energy loss in the evolution. As is shown
in Fig.~\ref{raa-phi-no-flow}, flow effects only slightly increase the quenching power of the medium in the
AMY-formalism. It is emphasized that for a realistic hydrodynamical calculation, the overall temperature of the
medium would drop not as fast  if collective flow was switched off and the medium itself would expand more
slowly in this case.

We point out that a further interesting quantity is $R_{AA}$ for neutral pions as a function of $p_T$ at
different centralities and away from midrapidity. The formalism as outlined in Section II. can be
straightforwardly extended to treat this case. Caveats are that only moderate deviations from midrapidity can be
allowed, because the nuclear parton distribution functions can be less exactly determined in the relevant region
\cite{Eskola:1998df} and the assumption of a thermalized medium essential for a hydrodynamical treatment is no
longer fulfilled far away from midrapidity. We therefore restrict our study to rapidities close to midrapidity
(maximum forward rapidity $y=2$).

\begin{figure}[htb]
\begin{center}
\includegraphics[width=8cm]{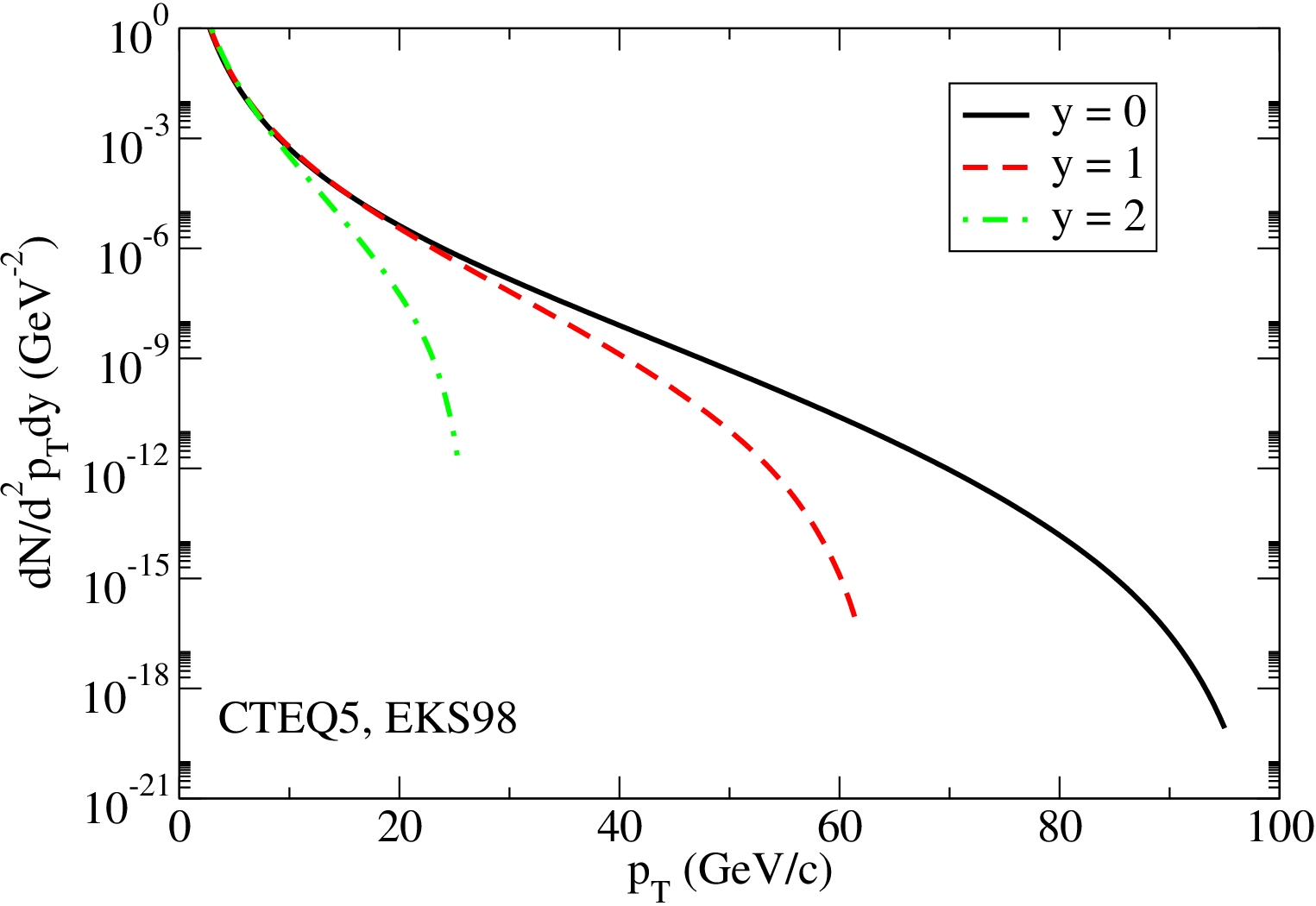}
\end{center}
\caption{(Color online) The jet (quark + anti-quark) transverse momentum distribution at different rapidities,
$b=2.4$~fm.}\label{jet-initial}
\end{figure}

\begin{figure}[htb]
\begin{center}
\includegraphics[width=8cm]{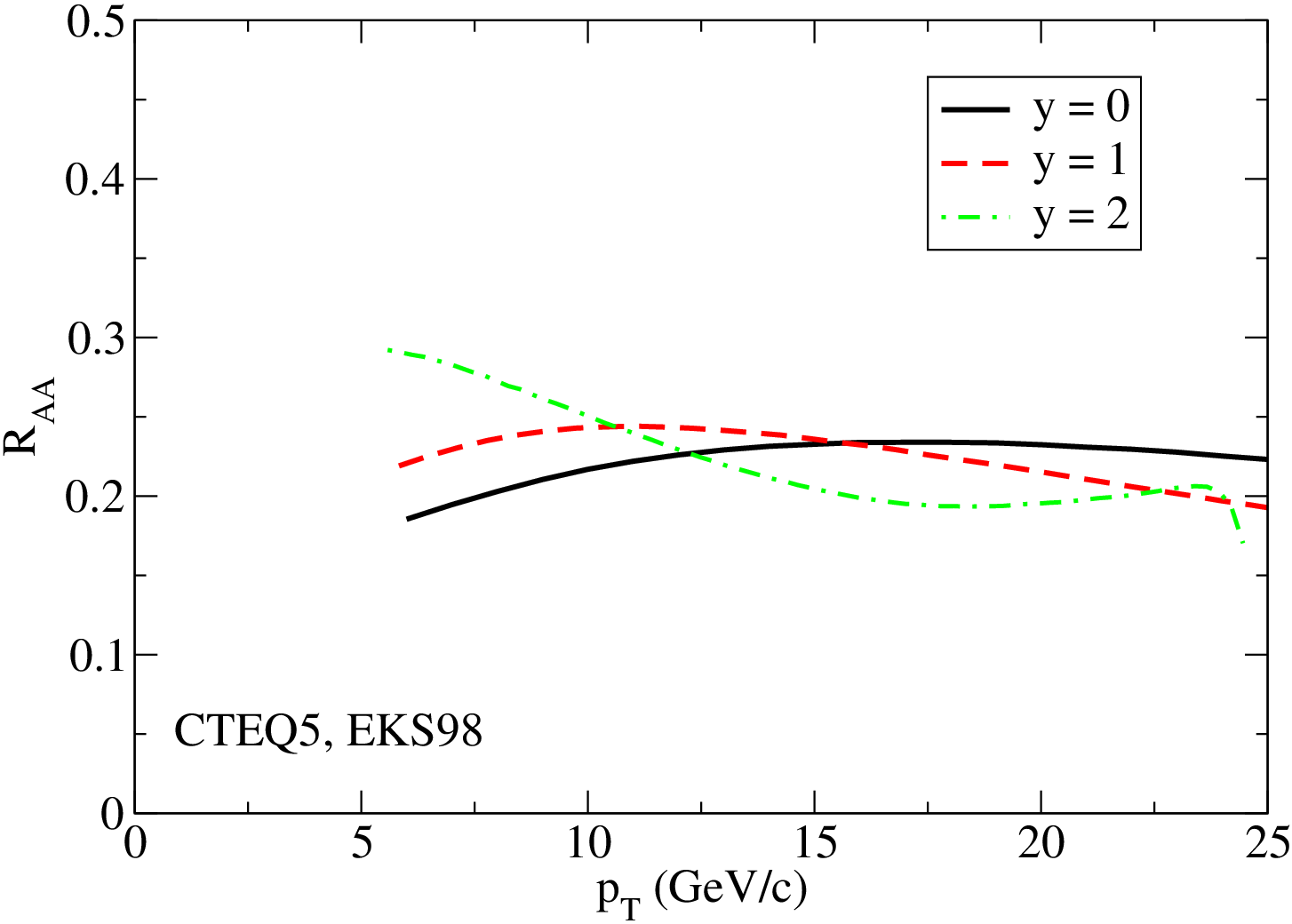}
\end{center}
\caption{(Color online) The neutral pion $R_{AA}$ at different rapidities,
$b=2.4$~fm.}\label{Raa-rapidity-central}
\end{figure}

At finite rapidity $y$ the energy of a highly-relativistic jet with a transverse momentum $p_T$ is given by
$E=p_T\cosh{y}$. The pions at a fixed $p_T$ have more energy and are the fragments of higher energetic partons
than the corresponding midrapidity pions. The initial jet distribution of quarks and anti-quarks is shown in
Fig.~\ref{jet-initial} for different rapidities, compare Eq.~(\ref{pdfeqn}). Note that the kinematical cut off
at $E=\sqrt{s_{NN}}/2 = 100~{\rm GeV}$ is reached at lower $p_T$ for finite $y$.

In Fig.~\ref{Raa-rapidity-central} we show $R_{AA}$ as a function of $p_T$ for central collisions ($0-5\%$,
$b=2.4$~fm) at mid and forward rapidity. It is interesting to notice that $R_{AA}$ behaves quite differently as
a function of $p_T$ at $y=2$ than at $y=0$. This is not only due to the different temperature profiles of the
hydrodynamical medium at forward rapidity but also strongly influenced by the different initial jet
distributions, see Fig.~\ref{jet-initial}.

\begin{figure}[htb]
\begin{center}
\includegraphics[width=8cm]{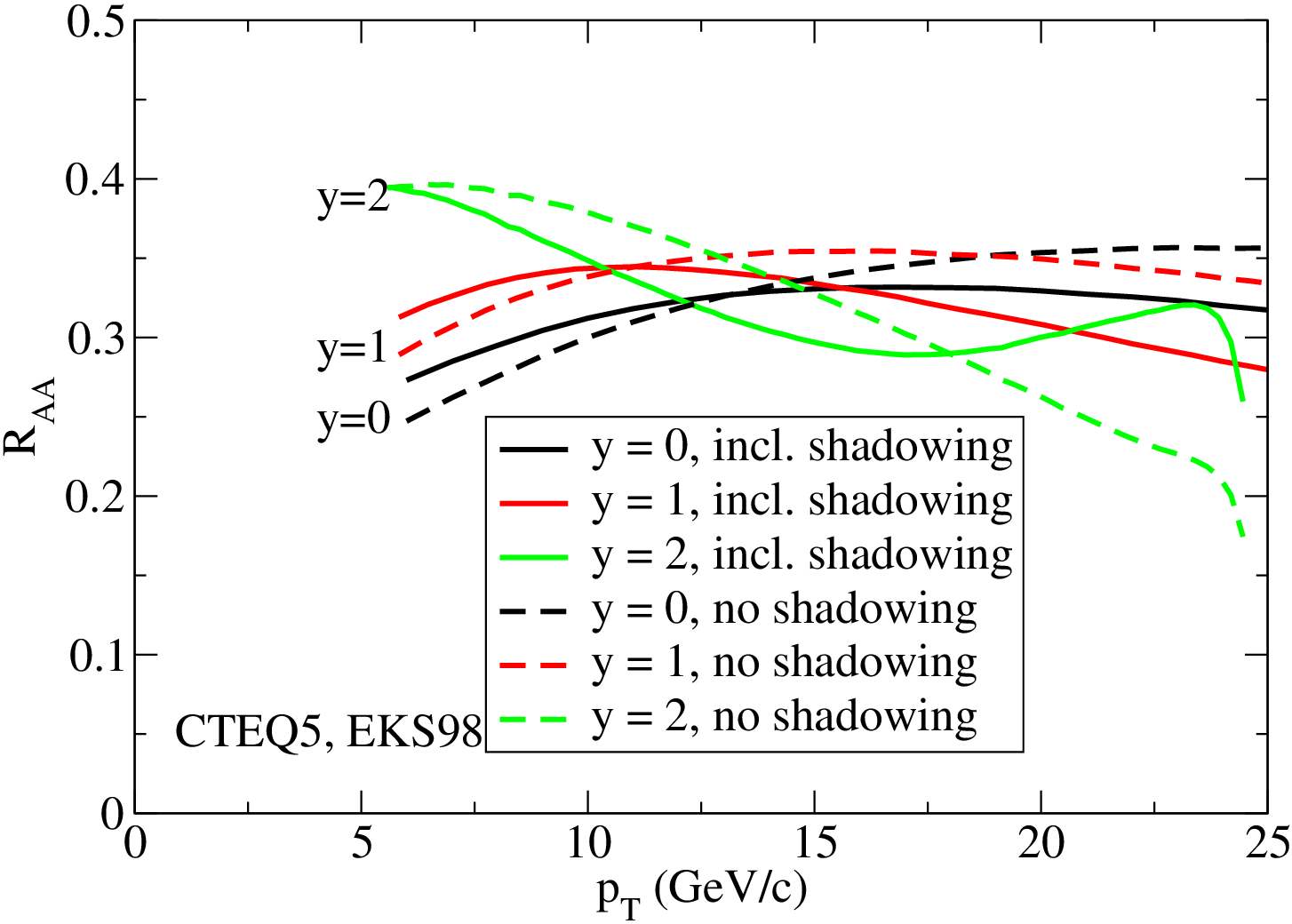}
\end{center}
\caption{(Color online) Comparing neutral pion $R_{AA}$ with and without nuclear shadowing effect at different
rapidities, $b=7.5$~fm.}\label{Raa-rapidity-non-central}
\end{figure}

To provide additional insight, we studied the same quantity averaged over $\phi$ for midcentral collisions with
an impact parameter of $b=7.5$~fm with and without nuclear shadowing effects taken into account in the parton
distribution functions utilized in Eq.~(\ref{pdfeqn}). Results are shown in Fig.~\ref{Raa-rapidity-non-central}.
It is interesting to notice that $R_{AA}$ is not monotonously increasing as a function of $p_T$. {The
midrapidity $R_{AA}$ is decreasing above $\sim 18$~GeV/c (with nuclear shadowing), the turning point for $y=1$
is at $\sim 9$~GeV/c (with nuclear shadowing). The values of $R_{AA}$ at $y=2$ decreases monotonically above
$\sim 6$~GeV/c in the case without nuclear shadowing and exhibits two turning points if shadowing is taken into
account.} We have also found that assuming a simple power law approximation for $dN/d^2p_T dy$ distributions for
all values of $p_T$ would lead to increased $R_{AA}$ at higher $p_T$ (comparison not shown). This demonstrates
that the overall decrease of $R_{AA}$ at higher $p_T$ is mainly due to the initial jet distribution according to
Eq.~(\ref{pdfeqn}) at high transverse momentum which decreases faster than an overall power law, see
Fig.~\ref{jet-initial}.

\begin{figure}[htb]
\begin{center}
\includegraphics[width=8cm]{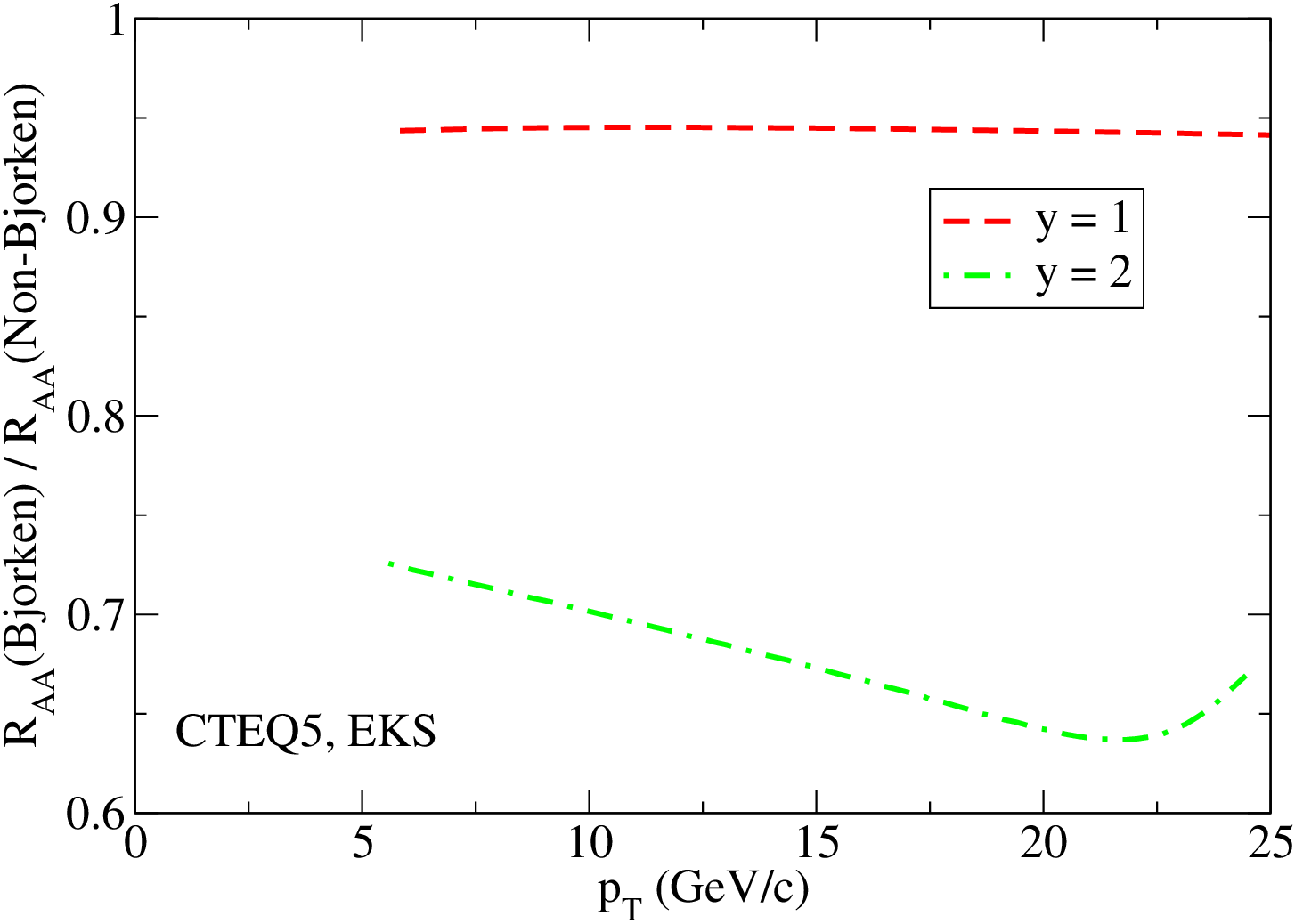}
\end{center}
\caption{(Color online) The ratio of the neutral pion $R_{AA}$ imposing a boost-invariant expansion to $R_{AA}$
as calculated from the 3D hydrodynamical (non-Bjorken) medium
 $b=7.5$~fm.}\label{raa-y-boost-ratio}
\end{figure}

One can also address the question how different the part of the medium is traversed by a jet which fragments
into a pion at forward rapidity in comparison to one which fragments at midrapidity. We compare the full 3D
hydrodynamical calculation to an effective 2D boost-invariant approach in which the 2D hydrodynamical solution
at midrapidity is assumed to also describe the transverse profile at forward rapidity. This corresponds
effectively to imposing {\it a posteriori} Bjorken expansion onto the non-Bjorken hydrodynamical evolution. We
study the ratio of $R_{\rm AA}$ by imposing a boost invariant expansion, and comparing with the fully 3D
non-Bjorken evolution. Fig.~\ref{raa-y-boost-ratio} shows a calculation at forward rapidities for non-central
collisions with a finite impact parameter of $b=7.5$~fm. This ratio is obviously not measurable, but is
interesting from a theoretical point of view. Its relatively strong deviations from $1$ at $y=2$ stem mainly
from the different transverse temperature profiles at forward rapidity in the non-Bjorken evolution whereas
these differences at $y=1$ are not significant. The fact that the ratio is rather flat in $p_T$ (it varies only
in the range of $0.7 \pm 0.05$ for $y=2$) indicates that the reduction of the quenching power of the medium in
the non-Bjorken case compared to the boost-invariant one is similar for partons over the full range of initial
jet energies probed in the collision. Therefore a measurement of  the absolute normalization of $R_{AA}$ at
midrapidity and forward rapidities might be useful in quantifying the deviations arising from the
simplifications made in boost invariant expansion models.

\subsection{Dependence on Nuclear Parton Distribution Functions}

\begin{figure}[htb]
\begin{center}
\includegraphics[width=8cm]{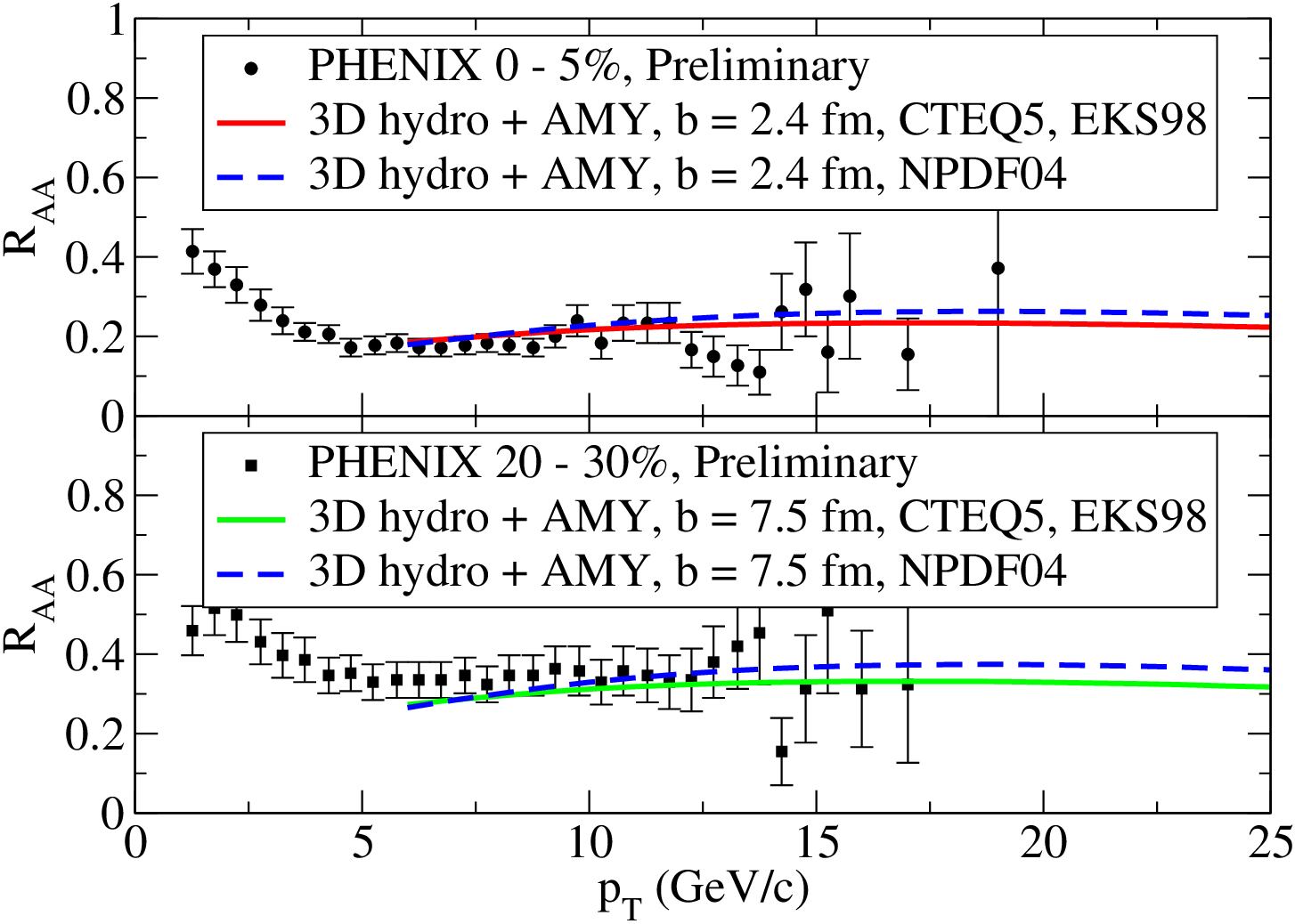}
\end{center}
\caption{(Color online) The neutral pion $R_{AA}$ at midrapidity in most central (upper panel) and midperipheral
(lower panel) Au+Au collisions compared with PHENIX data.
Different prescriptions of nuclear parton distribution functions are used for comparison. }
\label{npdf_RAA}
\end{figure}

\begin{figure}[htb]
\begin{center}
\includegraphics[width=8cm]{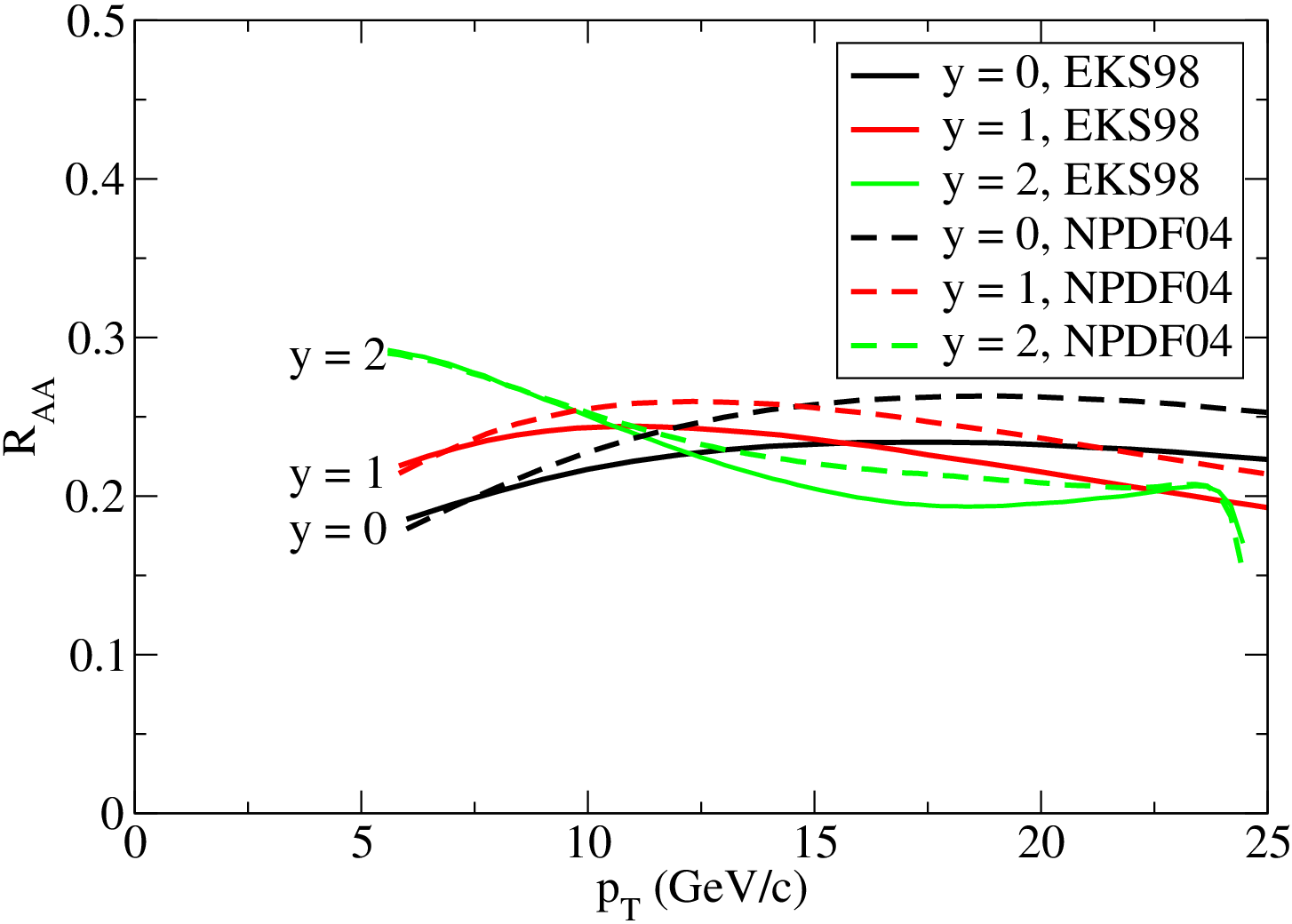}
\end{center}
\caption{(Color online) Comparing neutral pion $R_{AA}$ at different rapidities using different descriptions of
nuclear parton distribution functions, $b=2.4$~fm.}
\label{npdf_RAA_b2.4}
\end{figure}

As has been pointed out earlier, see e.g. \cite{Hirai:2004wq}, the determination of nuclear parton distribution functions (nuclear PDFs) from experimental data is ambiguous. These uncertainties can also influence the calculation of the nuclear modification factor at mid and forward rapidity at RHIC. We compare in this subsections results obtained with the nuclear parton distribution functions as determined
by NPDF04 \cite{Hirai:2004wq} with those that were employed so far in this work, namely EKS98.
We checked that the nucleon parton distributions which NPDF04 and EKS98 rely on, namely
MRST01 and CTEQ5, respectively, lead to almost the same prediction of the inclusive cross section for $\pi_0$ production in $p+p$ collisions, cmp. Figs. \ref{pp-cross-section} and \ref{pp-cross-section-y}. This should be expected since the determination of nucleon parton distribution functions has smaller uncertainties than those extended to nuclei.

Fig.~\ref{npdf_RAA} shows the neutral pion $R_{AA}$ at midrapidity in central and midperipheral Au+Au collisions
as obtained with the two different nuclear PDFs.
Differences due to the different nuclear PDFs appear especially at larger transverse momenta of the produced pions.
The same holds true for  $R_{AA}$ at forward rapidity, see Fig.~\ref{npdf_RAA_b2.4} for a comparison in
central collisions.

\begin{figure}[htb]
\begin{center}
\includegraphics[width=8cm]{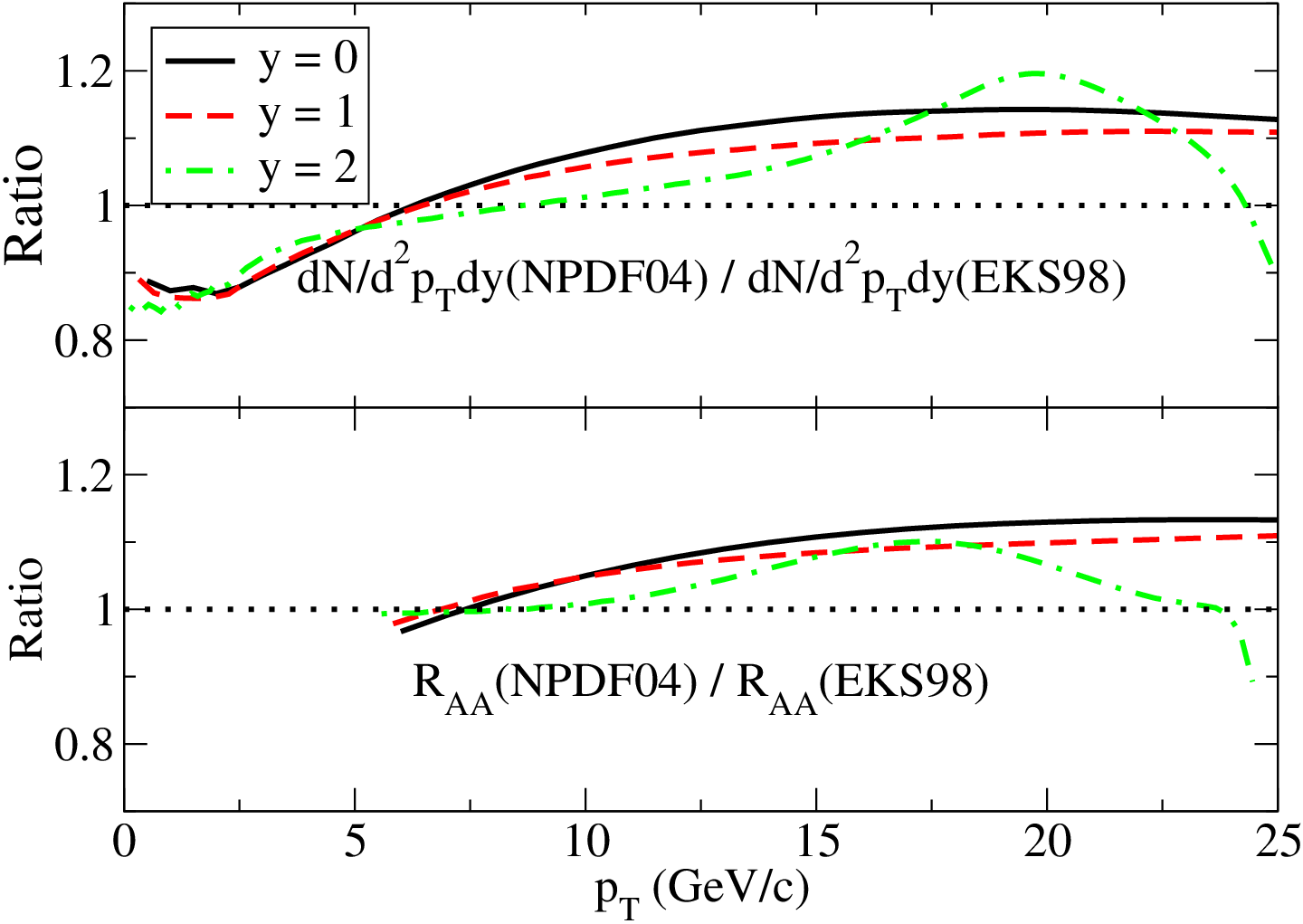}
\end{center}
\caption{(Color online)  The ratio (NPDF04/EKS98) for the initial quark + anti-quark jet distributions (upper)
and nuclear modification factors $R_{AA}$ (lower) using different descriptions of nuclear parton distribution functions, $b=2.4$~fm.}
\label{npdf_ratio_b2.4}
\end{figure}

It is possible to trace these differences in $R_{AA}$  back to differences in the initial jet distributions resulting mainly from the different
shadowing descriptions. We show in Fig.~\ref{npdf_ratio_b2.4} upper panel the ratio of the initial quark and anti-quark jet distributions as inferred from NPDF04 to EKS98. This translates - after jet-energy loss and fragmentation have been taken into account - into a similar behavior of the ratios of the nuclear suppression factor $R_{AA}$ in the two cases.
Differences in the initial distribution (mainly resulting from different nuclear shadowing) will therefore be reflected in
$R_{AA}$ at mid and forward rapidity and at different centralities.
Reduced sensitivity to differences in shadowing effects is expected if ratios of $R_{AA}$
are considered:  the ratio for $y=1$ (Fig. \ref{raa-y-boost-ratio}) is only sensitive on the $1\%$ level
to employing EKS98 or NPDF04 (comparison not shown), the sensitivity for $y=2$ is at most $4\%$.
What we find in this subsection clearly demonstrate that that $R_{AA}$ is not only sensitive to the employed jet quenching formalism but also to nuclear shadowing effects. The reason is that
-- even after energy loss and fragmentation -- $R_{AA}$ is sensitive to the initial jet distribution which
in turn vary within the uncertainties of the determination of nuclear shadowing. A further reduction of uncertainties in the determination of nuclear shadowing effects will make a more stringent test of jet quenching formalisms by $R_{AA}$ measurements.

%%%%%%%%%%%%%%%%%%%%%%%%%%%%%%%%%%%%%%%%%%%%%%%%%%%%%%%%%%%%%%%%%%%%%%%%%%%%%%%%%%%%%%%%%%%%%%%%%%%%%%%%%%%%%%%%%%%%%%%%
\section{Conclusions}
%%%%%%%%%%%%%%%%%%%%%%%%%%%%%%%%%%%%%%%%%%%%%%%%%%%%%%%%%%%%%%%%%%%%%%%%%%%%%%%%%%%%%%%%%%%%%%%%%%%%%%%%%%%%%%%%%%%%%%%%

In this paper, the jet energy loss was studied in the AMY-formalism using a 3D hydrodynamical evolution model
that has been shown to describe the bulk properties of matter created in heavy-ion collisions at RHIC.

We have evaluated the nuclear suppression factor $R_{AA}$ for neutral pions in central collisions as a function
of $p_T$ at mid and forward rapidity and have discussed how the azimuthal asymmetry of the medium in non-central
collisions allows to put stronger constraints on our understanding of jet energy loss by gluon radiation. Since
the jets probe different flow and temperature profiles in the asymmetric expansion depending on their initial
positions and emission angles, $R_{AA}$ is not only a function of $p_T$ but also of the azimuth in those
collisions.

The measured $R_{AA}$ as a function of $p_T$ in central and (averaged over the azimuth) in non-central
collisions at midrapidity is in good agreement with the model calculations. We also have provided calculations
of $R_{AA}$ as a function of $p_T$ and the azimuth without averaging that can be the basis of more stringent
experimental tests once further data become available. We furthermore studied $R_{AA}$ as a function of $p_T$ in
central and again averaged over the azimuth in non-central collisions at mid and forward rapidity and provided
arguments that a measurement of these dependences might not only be able to reveal more information about the
nuclear medium (as deviations from the assumption of boost invariance) but also provide a possibility to observe
nuclear shadowing effects  in the initial parton distribution function indirectly (assuming appropriate
experimental resolution).

We emphasize that the description of $R_{AA}$ (as a function of $p_T$,  the azimuthal angle and rapidity) alone
is not enough to prove the consistency of a specific energy loss mechanism with data, if assumptions about the
medium evolution can be freely adjusted. On the contrary, $R_{AA}$ will only provide stronger constraints on our
theoretical physical conjectures about jet energy loss in the nuclear medium if studied in a dynamical evolution
model which has been tested using soft observables.

%%%%%%%%%%%%%%%%%%%%%%%%%%%%%%%%%%%%%%%%%%%%%%%%%%%%%%%%%%%%%%%%%%%%%%%%%%%%%%%%%%%%%%%%%%%%%%%%%%%%%%%%%%%%%%%%%%%%%%%%
\section{Acknowledgments}
%%%%%%%%%%%%%%%%%%%%%%%%%%%%%%%%%%%%%%%%%%%%%%%%%%%%%%%%%%%%%%%%%%%%%%%%%%%%%%%%%%%%%%%%%%%%%%%%%%%%%%%%%%%%%%%%%%%%%%%%

It is a pleasure to thank S.~Jeon and T.~Renk for many discussions and comments. We thank W.~Vogelsang for
discussions. We thank Bryon Neufeld for computational assistance in extending our hydro-grid mapping routine to
incorporate collective flow. C.~G., G.-Y.~Q., J.~R., and S.~T. acknowledge financial support by the Natural
Sciences and Engineering Research Council of Canada. S. Bass acknowledges support by a grant from the U.S.
Department of Energy (DE-FG02-03ER41239-0).

%%%%%%%%%%%%%%%%%%%%%%%%%%%%%%%%%%%%%%%%%%%%%%%%%%%%%%%%%%%%%%%%%%%%%%%%%%%%%%%%%%%%%%%%%%%%%%%%%%%%%%%%%%%%%%%%%%%%%%%%


\begin{thebibliography}{99}

 \bibitem{Adcox:2001jp}
 K.~Adcox {\it et al.} [PHENIX Collaboration],
 % ``Suppression of hadrons with large transverse momentum in central Au + Au
 %collisions at s**(1/2)(N N) = 130-GeV,''
 Phys.\ Rev.\ Lett.\ {\bf 88}, 022301 (2002).
 %[arXiv:nucl-ex/0109003].
 %%CITATION = NUCL-EX 0109003;%%

\bibitem{Adler:2002xw}
 C.~Adler {\it et al.} [STAR Collaboration],
 %``Centrality dependence of high p(T) hadron suppression in Au + Au
 %collisions at s(NN)**(1/2) = 130-GeV,''
 Phys.\ Rev.\ Lett.\ {\bf 89} 202301 (2002).
 %[arXiv:nucl-ex/0206011].
 %%CITATION = NUCL-EX 0206011;%%


\bibitem{Gyulassy:1993hr}
 M.~Gyulassy and X.~Wang,
 %MULTIPLE COLLISIONS AND INDUCED GLUON BREMSSTRAHLUNG IN QCD.
 Nucl.\ Phys.\ {\bf 420} 583 (1994).


\bibitem{Baier:1996kr}
 R.~Baier, Y.~L.~Dokshitzer, A.~H.~Mueller, S.~Peigne and D.~Schiff,
 %``Radiative energy loss of high energy quarks and gluons in a
 %finite-volume quark-gluon plasma,''
 Nucl.\ Phys.\ B {\bf 483} 291 (1997).
 %[arXiv:hep-ph/9607355].
 %%CITATION = HEP-PH 9607355;%%

%\cite{Gyulassy:2000er}
\bibitem{Gyulassy:2000er}
 M.~Gyulassy, P.~Levai and I.~Vitev,
 %``Reaction operator approach to non-Abelian energy loss,''
 Nucl.\ Phys.\ B {\bf 594}, 371 (2001)
 %[arXiv:nucl-th/0006010].
 %%CITATION = NUPHA,B594,371;%%

%\cite{Kovner:2003zj}
\bibitem{Kovner:2003zj}
 A.~Kovner and U.~A.~Wiedemann,
 Review for Quark Gluon Plasma 3, Editors: R.C. Hwa and X.N. Wang,
 World Scientific, Singapore, 192 (2003),
 %``Gluon radiation and parton energy loss,''
 arXiv:hep-ph/0304151.
 %%CITATION = HEP-PH/0304151;%%

 %\cite{Zakharov:1996fv}
\bibitem{Zakharov:1996fv}
 B.~G.~Zakharov,
 %``Fully quantum treatment of the Landau-Pomeranchuk-Migdal effect in QED and
 %QCD,''
 JETP Lett.\ {\bf 63}, 952 (1996);
 %[arXiv:hep-ph/9607440].
 %%CITATION = JTPLA,63,952;%%
 %``Radiative energy loss of high energy quarks in finite-size nuclear matter
 %and quark-gluon plasma,''
 JETP Lett.\ {\bf 65}, 615 (1997);
 %[arXiv:hep-ph/9704255].
 %%CITATION = JTPLA,65,615;%%
 %``Transverse spectra of radiation processes in medium,''
 JETP Lett.\ {\bf 70}, 176 (1999);
 %[arXiv:hep-ph/9906536].
 %%CITATION = JTPLA,70,176;%%

\bibitem{Wang:2001if}
 X.~N.~Wang and X.~f.~Guo,
 %``Multiple parton scattering in nuclei: Parton energy loss,''
 Nucl.\ Phys.\ A {\bf 696} (2001) 788;
 %%CITATION = NUPHA,A696,788;%%
%\cite{Majumder:2004pt}
 A.~Majumder, E.~Wang and X.~N.~Wang,
 %``Modified dihadron fragmentation functions in hot and nuclear matter,''
 arXiv:nucl-th/0412061.
 %%CITATION = NUCL-TH/0412061;%%

%\cite{Arnold:2001ms}
\bibitem{Arnold:2001ms}
 P.~Arnold, G.~D.~Moore and L.~G.~Yaffe,
 %``Photon emission from ultrarelativistic plasmas,''
 JHEP {\bf 0111}, 057 (2001);
 %[arXiv:hep-ph/0109064].
 %%CITATION = JHEPA,0111,057;%%
 %``Photon emission from quark gluon plasma: Complete leading order results,''
 JHEP {\bf 0112}, 009 (2001);
 %%CITATION = JHEPA,0112,009;%%
  %``Photon and gluon emission in relativistic plasmas,''
 JHEP {\bf 0206}, 030 (2002).
 %[arXiv:hep-ph/0204343].
 %%CITATION = JHEPA,0206,030;%%

\bibitem{Dutt-Mazumder:2004xk}
 A.~K.~Dutt-Mazumder, J.~e.~Alam, P.~Roy and B.~Sinha,
 %``Stopping power of hot QCD plasma,''
 Phys.\ Rev.\ D {\bf 71} (2005) 094016.
 %[arXiv:hep-ph/0411015].
 %%CITATION = PHRVA,D71,094016;%%

%\cite{Adil:2006ei}
\bibitem{Adil:2006ei}
 A.~Adil, M.~Gyulassy, W.~A.~Horowitz and S.~Wicks,
 %``Collisional energy loss of non asymptotic jets in a QGP,''
 arXiv:nucl-th/0606010.
 %%CITATION = NUCL-TH/0606010;%%

%\cite{Mustafa:2003vh}
\bibitem{Mustafa:2003vh}
  M.~G.~Mustafa and M.~H.~Thoma,
  %``Quenching of hadron spectra due to the collisional energy loss of  partons
  %in the quark gluon plasma,''
  Acta Phys.\ Hung.\  A {\bf 22}, 93 (2005)
  %[arXiv:hep-ph/0311168].
  %%CITATION = APHUE,A22,93;%%

%\cite{Wang:2006qr}
\bibitem{Wang:2006qr}
 X.~N.~Wang,
 %``Interference effect in elastic parton energy loss in a finite medium,''
 arXiv:nucl-th/0604040.
 %%CITATION = NUCL-TH/0604040;%%

 %\cite{Dainese:2004te}
\bibitem{Dainese:2004te}
 A.~Dainese, C.~Loizides and G.~Paic,
 %``Leading-particle suppression in high energy nucleus nucleus collisions,''
 Eur.\ Phys.\ J.\ C {\bf 38} (2005) 461.
 %[arXiv:hep-ph/0406201].
 %%CITATION = EPHJA,C38,461;%%


%\cite{Majumder:2006we}
\bibitem{Majumder:2006we}
 A.~Majumder,
 %``Resolving the plasma profile via differential single inclusive
 %suppression,''
 Phys.\ Rev.\ C {\bf 75}, 021901 (2007).
 %[arXiv:nucl-th/0608043].
 %%CITATION = PHRVA,C75,021901;%%

%\cite{Hirano:2003hy}
\bibitem{Hirano:2003hy}
 T.~Hirano and Y.~Nara,
 %``Energy loss in high energy heavy ion collisions from the hydro+jet
 %model,''
 Phys.\ Rev.\ C {\bf 66}, 041901 (2002).
 %[arXiv:hep-ph/0208029].
 %%CITATION = PHRVA,C66,041901;%%

%\cite{Renk:2005ta}
\bibitem{Renk:2005ta}
 T.~Renk and J.~Ruppert,
 %``Flow dependence of high p(T) parton energy loss in heavy-ion collisions,''
 Phys.\ Rev.\ C {\bf 72}, 044901 (2005).
 %[arXiv:hep-ph/0507075].
 %%CITATION = PHRVA,C72,044901;%%

%\cite{Renk:2004yv}
\bibitem{Renk:2004yv}
 T.~Renk,
 %``A dynamical model for the spacetime evolution of heavy-ion collisions at
 %RHIC,''
 Phys.\ Rev.\ C {\bf 70} (2004) 021903.
 %[arXiv:hep-ph/0404140].
 %%CITATION = PHRVA,C70,021903;%%

%\cite{Salgado:2003gb}
\bibitem{Salgado:2003gb}
 C.~A.~Salgado and U.~A.~Wiedemann,
 %``Calculating quenching weights,''
 Phys.\ Rev.\ D {\bf 68}, 014008 (2003).
 %[arXiv:hep-ph/0302184].
 %%CITATION = PHRVA,D68,014008;%%
%\cite{Renk:2006sx}

\bibitem{RenkEskola}
 T.~Renk and K.~J.~Eskola,
 %``Prospects of medium tomography using back-to-back hadron correlations,''
 arXiv:hep-ph/0610059.

%\cite{Nonaka:2006yn}
\bibitem{Nonaka:2006yn}
 C.~Nonaka and S.~A.~Bass,
 %``Space-time evolution of bulk QCD matter,''
 Phys.\ Rev.\ C {\bf 75}, 014902 (2007).
 %[arXiv:nucl-th/0607018].
 %%CITATION = PHRVA,C75,014902;%%

\bibitem{Renk:2006sx}
 T.~Renk, J.~Ruppert, C.~Nonaka and S.~A.~Bass,
 %``Jet-quenching in a 3D hydrodynamic medium,''
 arXiv:nucl-th/0611027.
 %%CITATION = NUCL-TH/0611027;%%


 %\cite{Bass:2007em}
\bibitem{Bass:2007em}
 S.~A.~Bass, T.~Renk, J.~Ruppert and C.~Nonaka,
 %``Hard and soft probe - medium interactions in a 3D hydro+micro approach at
 %RHIC,''
 arXiv:nucl-th/0702079.
 %%CITATION = NUCL-TH/0702079;%%


\bibitem{Majumder2}
 A.~Majumder, C.~Nonaka and S.~A.~Bass,
 %``Jet modification in three dimensional fluid dynamics at next-to-leading
 %twist,''
 arXiv:nucl-th/0703019.


%\cite{Renk:2006pw}
\bibitem{Renk:2006pw}
 T.~Renk,
 %``Mach cones and dijets: Jet quenching and fireball expansion dynamics,''
 arXiv:hep-ph/0608333.
 %%CITATION = HEP-PH/0608333;%%



%\cite{Adler:2}
\bibitem{Adler:2}
 S.~S.~Adler {\it et al.} [PHENIX Collaboration],
 %``A detailed study of high-p(T) neutral pion suppression and azimuthal
 %anisotropy in Au + Au collisions at s(NN)**(1/2) = 200-GeV,''
 arXiv:nucl-ex/0611007.
 %%CITATION = NUCL-EX/0611007;%%

\bibitem{RenkRuppert1}
 T.~Renk and J.~Ruppert,
 %``Mach cones in an evolving medium,''
 Phys.\ Rev.\ C {\bf 73} (2006) 011901
 %%CITATION = PHRVA,C73,011901;%%

\bibitem{RenkRuppert2}
 T.~Renk and J.~Ruppert,
 %``Three-particle azimuthal correlations and Mach shocks,''
 arXiv:hep-ph/0702102.
 %%CITATION = HEP-PH/0702102;%%

%\cite{Bjorken:1982qr}
\bibitem{Bjorken:1982qr}
 J.~D.~Bjorken,
 %``Highly Relativistic Nucleus-Nucleus Collisions: The Central Rapidity
 %Region,''
 Phys.\ Rev.\ D {\bf 27}, 140 (1983).
 %%CITATION = PHRVA,D27,140;%%

%\cite{Clare:1986qj}
\bibitem{Clare:1986qj}
 R.~B.~Clare and D.~Strottman,
 %``Relativistic Hydrodynamics and heavy ion reactions,''
 Phys.\ Rept.\ {\bf 141}, 177 (1986).
 %%CITATION = PRPLC,141,177;%%

%\cite{Dumitru:1998es}
\bibitem{Dumitru:1998es}
 A.~Dumitru and D.~H.~Rischke,
 %``Collective dynamics in highly relativistic heavy-ion collisions,''
 Phys.\ Rev.\ C {\bf 59}, 354 (1999).
 %[arXiv:nucl-th/9806003].
 %%CITATION = PHRVA,C59,354;%%

\bibitem{Kolb:2001qz}
 P.~F.~Kolb, U.~W.~Heinz, P.~Huovinen, K.~J.~Eskola and K.~Tuominen,
 %``Centrality dependence of multiplicity, transverse energy, and elliptic
 %flow from hydrodynamics,''
 Nucl.\ Phys.\ A {\bf 696} (2001) 197.
 %[arXiv:hep-ph/0103234].
 %%CITATION = NUPHA,A696,197;%%

%\cite{Fries:2003kq}
\bibitem{Fries:2003kq}
 R.~J.~Fries, B.~Muller, C.~Nonaka and S.~A.~Bass,
 %``Hadron production in heavy ion collisions: Fragmentation and recombination
 %from a dense parton phase,''
 Phys.\ Rev.\ C {\bf 68}, 044902 (2003)
 %[arXiv:nucl-th/0306027].
 %%CITATION = PHRVA,C68,044902;%%

%\bibitem{Turbide}
%S. Turbide, C. Gale, E. Frodermann, J. Ruppert, U. Heinz, in preparation.

%\cite{DeJager:1974dg}
\bibitem{DeJager:1974dg}
  C.~W.~De Jager, H.~De Vries and C.~De Vries,
  %``Nuclear charge and magnetization density distribution parameters from
  %elastic electron scattering,''
  Atom.\ Data Nucl.\ Data Tabl.\  {\bf 14}, 479 (1974).
  %%CITATION = ADNDA,14,479;%%

%\cite{Lai:1999wy}
\bibitem{Lai:1999wy}
 H.~L.~Lai {\it et al.} [CTEQ Collaboration],
 %``Global {QCD} analysis of parton structure of the nucleon: CTEQ5 parton
 %distributions,''
 Eur.\ Phys.\ J.\ C {\bf 12}, 375 (2000).
 % [arXiv:hep-ph/9903282].
 %%CITATION = EPHJA,C12,375;%%
%\cite{Eskola:1998df}
\bibitem{Eskola:1998df}
 K.~J.~Eskola, V.~J.~Kolhinen and C.~A.~Salgado,
 %``The scale dependent nuclear effects in parton distributions for practical
 %applications,''
 Eur.\ Phys.\ J.\ C {\bf 9}, 61 (1999).
 %[arXiv:hep-ph/9807297].
 %%CITATION = EPHJA,C9,61;%%
%\cite{Eskola:2005ue}

\bibitem{Eskola:2005ue}
  K.~J.~Eskola, H.~Honkanen, H.~Niemi, P.~V.~Ruuskanen and S.~S.~Rasanen,
  %``RHIC-tested predictions for low-p(T) and high-p(T) hadron spectra in
  %nearly central Pb + Pb collisions at the LHC,''
  Phys.\ Rev.\  C {\bf 72}, 044904 (2005)
  %[arXiv:hep-ph/0506049].
  %%CITATION = PHRVA,C72,044904;%%

%\cite{Jager:2002xm}
\bibitem{Jager:2002xm}
  B.~J\"ager, A.~Sch\"afer, M.~Stratmann and W.~Vogelsang,
  %``Next-to-leading order QCD corrections to high-p(T) pion production in
  %longitudinally polarized p p collisions,''
  Phys.\ Rev.\  D {\bf 67}, 054005 (2003)
  %[arXiv:hep-ph/0211007].
  %%CITATION = PHRVA,D67,054005;%%

 %\cite{Barnafoldi:2000dy}
\bibitem{Barnafoldi:2000dy}
 G.~G.~Barnafoldi, G.~I.~Fai, P.~Levai, G.~Papp and Y.~Zhang,
 %``Jets and produced particles in p p collisions from SPS to RHIC energies
 %for nuclear applications,''
 J.\ Phys.\ G {\bf 27}, 1767 (2001).
 %[arXiv:nucl-th/0004066].
 %%CITATION = JPHGB,G27,1767;%%

%\cite{Adler:2006wg}
\bibitem{Adler:2006wg}
 S.~S.~Adler {\it et al.} [PHENIX Collaboration],
 %``Centrality dependence of pi0 and eta production at large transverse
 %momentum in s(NN)**(1/2) = 200-GeV d + Au collisions,''
 arXiv:nucl-ex/0610036.
 %%CITATION = NUCL-EX/0610036;%%

%\cite{Jeon:2003gi}
\bibitem{Jeon:2003gi}
 S.~Jeon and G.~D.~Moore,
 %``Energy loss of leading partons in a thermal QCD medium,''
 Phys.\ Rev.\ C {\bf 71}, 034901 (2005)
 %[arXiv:hep-ph/0309332].
 %%CITATION = PHRVA,C71,034901;%%

%\cite{Turbide:2005fk}
\bibitem{Turbide:2005fk}
 S.~Turbide, C.~Gale, S.~Jeon and G.~D.~Moore,
 %``Energy loss of leading hadrons and direct photon production in evolving
 %quark-gluon plasma,''
 Phys.\ Rev.\ C {\bf 72}, 014906 (2005).
 %[arXiv:hep-ph/0502248].
 %%CITATION = PHRVA,C72,014906;%%

%\cite{Aurenche:2002pd}
\bibitem{Aurenche:2002pd}
  P.~Aurenche, F.~Gelis and H.~Zaraket,
  %``A simple sum rule for the thermal gluon spectral function and
  %applications,''
  JHEP {\bf 0205}, 043 (2002)
  %[arXiv:hep-ph/0204146].
  %%CITATION = JHEPA,0205,043;%%

%\cite{Kniehl:2000fe}
\bibitem{Kniehl:2000fe}
  B.~A.~Kniehl, G.~Kramer and B.~Potter,
  %``Fragmentation functions for pions, kaons, and protons at  next-to-leading
  %order,''
  Nucl.\ Phys.\  B {\bf 582}, 514 (2000)
  %[arXiv:hep-ph/0010289].
  %%CITATION = NUPHA,B582,514;%%

%%\cite{Binnewies:1995pt}
%\bibitem{Binnewies:1995pt}
%  J.~Binnewies, B.~A.~Kniehl and G.~Kramer,
%  %``Pion And Kaon Production In E+ E- And E P Collisions At Next-To-Leading
%  %Order,''
%  Phys.\ Rev.\  D {\bf 52}, 4947 (1995)
%  [arXiv:hep-ph/9503464].
%  %%CITATION = PHRVA,D52,4947;%%
%\cite{Adams:2006uz}

\bibitem{phenix-pi0-pp}
http://www.phenix.bnl.gov/WWW/plots/show\_plot.php ?editkey=p0439

%\cite{Martin:2002dr}
\bibitem{Martin:2002dr}
  A.~D.~Martin, R.~G.~Roberts, W.~J.~Stirling and R.~S.~Thorne,
  %``NNLO global parton analysis,''
  Phys.\ Lett.\  B {\bf 531}, 216 (2002)
  [arXiv:hep-ph/0201127].
  %%CITATION = PHLTA,B531,216;%%

\bibitem{Adams:2006uz}
  J.~Adams {\it et al.}  [STAR Collaboration],
  %``Forward neutral pion production in p+p and d+Au collisions at  s(NN)**(1/2)
  %= 200-GeV,''
  Phys.\ Rev.\ Lett.\  {\bf 97}, 152302 (2006)
  %[arXiv:nucl-ex/0602011].
  %%CITATION = PRLTA,97,152302;%%

%\bibitem{phenix-pi0-AA}
%http://www.phenix.bnl.gov/WWW/plots/show\_plot.php ?editkey=p0386

%\cite{Hirai:2004wq}
\bibitem{Hirai:2004wq}
  M.~Hirai, S.~Kumano and T.~H.~Nagai,
  %``Nuclear parton distribution functions and their uncertainties,''
  Phys.\ Rev.\  C {\bf 70}, 044905 (2004)
  [arXiv:hep-ph/0404093].
  %%CITATION = PHRVA,C70,044905;%%



\end{thebibliography}
\end{document}